\documentclass{lmcs}
\pdfoutput=1

\usepackage{lastpage}
\lmcsdoi{18}{3}{8}
\lmcsheading{}{\pageref{LastPage}}{}{}%
{Mar.~31,~2021}{Jul.~28,~2022}{}

\usepackage{amssymb}
\usepackage{hyperref}

\usepackage{tikz-cd}
\usepackage{yfonts}
\usepackage[utf8]{inputenc}

\keywords{computable analysis, discontinuous functions, envelopes, complete lattices}

\newcommand{\dom}{\operatorname{dom}}
\newcommand{\id}{\operatorname{id}}
\newcommand{\R}{\mathbb{R}}
\newcommand{\C}{\mathbb{C}}
\newcommand{\Q}{\mathbb{Q}}
\newcommand{\N}{\mathbb{N}}

\renewcommand{\O}{\mathcal{O}}
\newcommand{\A}{\mathcal{A}}
\newcommand{\V}{\mathcal{V}}
\newcommand{\E}{\mathfrak{E}}
\newcommand{\K}{\mathcal{K}}

\newcommand{\Kb}{\mathcal{K}_{\bot}}

\newcommand{\clos}{\operatorname{cl}}
\newcommand{\intr}[1]{#1^{\circ}}
\newcommand{\ucl}[1]{\left\uparrow{#1}\right.}

\newlength\arrowheight

\newcommand{\Set}[2]{\left\{#1 \mid #2\right\}}

\newcommand{\Sheaf}[2]{\mathcal{C}_{#1}(#2)}

\newcommand{\lift}[2]{\left(\raisebox{-0.2em}{$#1$} \hspace{-0.25em} {\big \backslash} \hspace{-0.2em} \raisebox{0.25em}{$#2$}\right)}

\newcommand{\starred}[1]{#1^\star}

\title[Uniform Envelopes]{Uniform Envelopes}
\author[E.~Neumann]{Eike Neumann}
\address{Swansea University, Swansea, UK}
\email{neumaef1@gmail.com}

\begin{document}
\begin{abstract}
	In the author's PhD thesis (2019) universal envelopes were introduced as a tool for studying the continuously obtainable information 
	on discontinuous functions with the aim of extending the scope of traditional computable analysis.

	To any function $f \colon X \to Y$ between $\operatorname{qcb}_0$-spaces one can assign a so-called universal envelope which, in a well-defined sense, encodes all continuously obtainable information on the function.
	A universal envelope consists of two continuous functions
	$F \colon X \to L$
	and 
	$\xi_L \colon Y \to L$
	with values in a $\Sigma$-split injective space $L$.
	Any continuous function with values in a $\Sigma$-split injective space whose composition with the original function is again continuous 
	factors through the universal envelope.
	This leaves open the question how to obtain such a factorisation for a given function.
	
	By an observation which essentially goes back to Scott (1972) any continuous function with values in a $\Sigma$-split injective space admits a greatest continuous ``right extension'' along $\xi_L$, which yields an explicit solution.
	However, this extension is not in general computable from the function.
	It follows from a result of Escard\'o (1998) that this extension depends continuously on the function if and only if $\xi_L$ is a proper map in the sense of Hofmann and Lawson (1984).

	In order to solve this problem in greater generality we propose the notion of uniform envelopes.
	A uniform envelope is additionally endowed with a map 
	$u_L \colon L \to \O^2(Y)$
	that is compatible with the multiplication of the double powerspace monad $\O^2$ in a certain sense.
	This yields for every continuous map with values in a $\Sigma$-split injective space
	a choice of uniformly computable extension.
	Under a suitable condition which we call uniform universality, 
	this extension yields a uniformly computable solution for the above factorisation problem.
	
	Uniform envelopes can be endowed with a composition operation.
	We establish criteria that ensure that the composition of two uniformly universal envelopes is again uniformly universal.
	These criteria admit a partial converse and we provide evidence that they cannot be easily improved in general.
	
	We show that a $\operatorname{qcb}_0$-space $Y$ has the property that all functions with values in $Y$ 
	admit a uniformly universal envelope if and only if $Y$ is Noetherian.
	Dually, a space $X$ has the property that all functions with domain $X$ admit a uniformly universal envelope 
	if and only if $X$ contains no infinite compact sets.
	
	Not every function admits a uniformly universal uniform envelope.
	We can however assign to every function a canonical envelope that is in some sense as close as possible to a uniform envelope.
	We obtain a composition theorem similar to the uniform case.
\end{abstract}
\maketitle

\section{Introduction, Background, and Motivation}

Computable analysis is the extension of the theory of computation from discrete to continuous data.
As such, it offers itself as a rigorous foundation for numerical computation.
The main theorem of computable analysis asserts that all computable functions are continuous.
Yet, discontinuous functions play an important role in numerical analysis, 
both as solution operators of important computational problems,
such as algebraic or differential equations,
and as building blocks for stable algorithms,
such as testing for inequality or matrix diagonalisation. 
In \cite{NeumannPhD} \emph{continuous envelopes} were introduced as a way of systematically studying the amount of continuously available information on a discontinuous function with the two-fold purpose of enabling the computation of partial information on problems with discontinuous solution operators and of facilitating the use of discontinuous operations within rigorous numerical software.
An envelope of a function 
$f \colon X \to Y$ 
between effective $T_0$ spaces is a pair of continuous maps
$F \colon X \to L$ 
and 
$\xi_L \colon Y \to L$ 
with values in an effective injective space $L$  
such that 
$\xi_L \circ f \geq F$ 
with respect to the specialisation order.

Before we proceed, let us review these notions.
We assume familiarity with the basic definitions and results of computable analysis, as for instance presented in \cite{PaulyRepresented}.
We will also make heavy use of the definitions and results of \cite{NeumannPhD}.
A detailed introduction that covers everything we need is given in \cite[Chapter 2]{NeumannPhD}.
We denote Sierpinski space by $\Sigma$.
For a represented space $X$ we let $\O(X) = \Sigma^X$ denote the space of opens.
We let $\K(X) \subseteq \O^2(X)$ denote the space of compacts and $\V(X) \subseteq \O^2(X)$ denote the space of overts.
We call $X$ an \emph{effective $T_0$ space} if the natural map 
$\nu_X \colon X \to \O^2(X)$
is an embedding.
We call $X$ a \emph{computable $T_0$ space} if $\nu_X$ is a computable embedding.
All effective $T_0$ spaces are endowed with the final topology of their representation.
In this sense, the space $\O(X)$ carries the Scott topology. 
This has the following somewhat intriguing consequence:
A continuous map $f \colon X \to Y$ induces a canonical continuous map $f^* \colon \O(Y) \to \O(X)$.
Any open, not necessarily continuous, map $f \colon X \to Y$ induces a canonical continuous map $f_* \colon \O(X) \to \O(Y)$.
Any effective $T_0$ space can be made into a partially ordered set by virtue of the specialisation order,
where $x \leq y$ if and only if for all $U \in \O(X)$ with $x \in U$ we have $y \in U$.
Unless otherwise stated, the symbol ``$\leq$'' always refers to the specialisation order on a space.

An 
\emph{effective injective space} 
is a pair of effective $T_0$ spaces 
$(L, A_L)$
together with a continuous section-retraction pair 
\begin{center}
\begin{tikzcd}
    \arrow[loop left]{l}{\id_L} L \arrow[swap,shift right = 1.2]{r}{s_L} & \arrow[swap,shift right = 1.2]{l}{r_L} \O(A_L).
\end{tikzcd}
\end{center}
If $L$ and $A_L$ are computable $T_0$ spaces and $s_L$ and $r_L$ are computable then we call 
$(L,A_L,s_L,r_L)$ 
a \emph{computable injective space}.
This data induces a left inverse 
$\rho_L \colon \O^2(L) \to L$
of the natural embedding 
$\nu_L \colon L \to \O^2(L)$
by letting 
$
    \rho_L = r \circ \nu_{A_L}^* \circ s^{**}
$.
Any effective injective space is simultaneously a $\V$-algebra and a $\K$-algebra.
It is easy to see that the structure maps are necessarily given by join and meet in the specialisation order respectively.
This implies that any effective injective space is a complete lattice with respect to its specialisation order.
More generally, we call any effective $T_0$ space $L$ which is simultaneously an effective $\V$-algebra and an effective $\K$-algebra an \emph{effective complete lattice}.

The retracts of spaces of the form $\O(A_L)$ for some represented space $A_L$ are the injective objects in the category of effective $T_0$ spaces relative to the class of $\Sigma$-split embeddings.
A $\Sigma$-split embedding is a map $e \colon X \to Y$ such that the induced map $e^* \colon \O(Y) \to \O(X)$ has a continuous section $s \colon \O(X) \to \O(Y)$.
The importance of this class of maps was probably first noticed by P. Taylor \cite{ASDSubspaces}.
For a given continuous function $f \colon X \to L$ and a given $\Sigma$-split embedding $e \colon X \to Y$ with a choice of section $s \colon \O(X) \to \O(Y)$ of $e^*$ the data $(L,A_L,s_L,r_L)$ can be used to compute the extension 
$\rho_L \circ f^{**} \circ s^* \circ \nu_Y$
of $f$ along $e$.
Thus, an effective injective space constitutes an effective injective object with a choice of extension of type
$
    L^X \to L^Y
$
for all $e \colon X \to Y$ with a chosen section $s \colon \O(X) \to \O(Y)$.

An \emph{effective continuous lattice} is an effective injective space $(L, A_L)$ where $A_L$ embeds into $\N$.
By a theorem of Smyth \cite{Smyth} we recover the classical definition of effective continuous lattice, see also \cite[Section 2.5]{NeumannPhD}.
From this point of view effective injective spaces are a very straightforward generalisation of effective continuous domains -- the injective objects in the category of topological spaces relative to subspace embeddings.
For a more thorough discussion of this, see \cite[Section 3.1]{NeumannPhD}.

The class of effective injective spaces is in a sense much larger than the class of effective continuous domains. 
The space $\O(\N^\N)$ is a computable injective space which is not locally compact and hence not a continuous domain.
The space $\O^2(\N^\N)$ is a quasi-Polish \cite{DeBrechtQuasiPolish} computable injective space which is not a continuous domain. 
Crucially, the space $\O^2(X)$ is an effective injective space for every effective $T_0$ space.

Effective and computable injective spaces have excellent closure properties: they are closed under finite products, exponentiation, and retracts
\cite[pp. 56 -- 57]{NeumannPhD}.

An 
\emph{effective approximation space} 
is an effective injective space $L$ together with a continuous map 
$\xi_L \colon Y \to L$.
Again, a 
\emph{computable approximation space}
is one where $L$ and $\xi_L$ are computable.

Let $f \colon X \to Y$ be a function between effective $T_0$ spaces.
As already partially defined above, an envelope of $f$ is an effective approximation space $\xi_L \colon Y \to L$
together with a continuous map $F \colon X \to L$ satisfying 
$F \leq \xi_L \circ f$.
We will usually simply write $F \colon X \to L$ for an envelope,
leaving the rest of the data implicit.

If $f \colon X \to Y$ is any function and $\xi_L \colon Y \to L$ is an approximation space for $Y$ 
then there exists a best continuous approximation $F \colon X \to L$,
in the sense that $F(x) \leq \xi_L \circ f(x)$ for all $x \in X$,
and if a continuous map $G \colon X \to L$ satisfies $G(x) \leq \xi_L \circ f(x)$ for all $x \in X$,
then $G(x) \leq F(x)$ for all $x \in X$.
We call this $F$ the \emph{principal $L$-envelope} of $f$.

Let $F \colon X \to L$ and $G \colon X \to M$ be envelopes of $f \colon X \to Y$.
We say that $F$ \emph{tightens} $G$ if there exists a continuous map $\Phi \colon L \to M$
satisfying $\Phi \circ \xi_L \leq \xi_M$ and $\Phi \circ F \geq G$.

An envelope $F \colon X \to L$ of $f \colon X \to Y$ is called \emph{universal} if it tightens all 
envelopes of $f$.
Any function $f \colon X \to Y$ between effective $T_0$ spaces has a universal envelope \cite[Theorem 4.8]{NeumannPhD}.


The next two results make the idea precise that an envelope encodes all continuously obtainable information on a function:

\begin{propC}[{\cite[Theorem 4.31]{NeumannPhD}}]\label{Proposition: extending global probe, non-uniform}
    Let $f \colon X \to Y$ be a function between effective $T_0$ spaces.
    Let $F \colon X \to L$ be a universal envelope of $f$.
    Let $\varphi \colon Y \to Z$ be a continuous map.
    Then $\varphi \circ f$ is continuous if and only if there exists a
    continuous map 
    $\widetilde{\nu_Z \circ \varphi} \colon L \to \O^2(Z)$
    satisfying 
    $\widetilde{\nu_Z \circ \varphi} \circ \xi_L \leq \nu_Z \circ \varphi$
    and 
    $\widetilde{\nu_Z \circ \varphi} \circ F = \nu_Z \circ \varphi \circ f$.
\end{propC}

\begin{propC}[{\cite[Theorem 4.34, Proposition 4.35]{NeumannPhD}}]\label{Proposition: extending probe, non-uniform}
    Let $f \colon X \to Y$ be a function between effective $T_0$ spaces.
    Let $F \colon X \to L$ be a universal envelope of $f$.
    Let 
    $\alpha \colon \widetilde{X} \to X$ 
    and 
    $\beta \colon \widetilde{X} \times Y \to Z$
    be continuous functions.
    Assume that there exists an embedding $i \colon Z \to M$
    into an effective continuous lattice.
    Then the following are equivalent:
    \begin{enumerate}
        \item 
            Every point of the form 
            $(\widetilde{x}, \alpha(\widetilde{x})) \in \widetilde{X} \times X$
            is a point of continuity of the map 
            $
                \widetilde{X} \times X \to Z,
                \; 
                (\widetilde{x}, x) \mapsto \beta(\widetilde{x}, f (x)).
            $
        \item 
            There exists a continuous map
            $
                \widetilde{i \circ \beta} \colon \widetilde{X} \times L \to M
            $
            satisfying 
            $
                \widetilde{i \circ \beta}(\widetilde{x}, \xi_L(y)) \leq i \circ \beta(\widetilde{x}, y)
            $ 
            and 
            $
                \widetilde{i \circ \beta}(\widetilde{x}, F \circ \alpha(\widetilde{x}))
                = 
                i \circ \beta(\widetilde{x}, f \circ \alpha(\widetilde{x})).
            $
    \end{enumerate}
\end{propC}

Propositions
\ref{Proposition: extending global probe, non-uniform}
and 
\ref{Proposition: extending probe, non-uniform}
are somewhat unsatisfactory.
They assert the existence of functions with certain properties but do not 
yield an effective method that allows one to find such functions.

The issue stems from the non-constructivity of tightening relation:
an envelope $F$ tightens an envelope $G$ if there exists a map 
$\Phi \colon L \to M$
satisfying 
$\Phi \circ \xi_L \leq \xi_M$ 
and 
$\Phi \circ F \geq G$.
But the definition does not tell us how to find such a map.
By an observation of Escard\'o \cite{EscardoInjective} which essentially goes back to Scott \cite{ScottLattices} there always exists 
a greatest continuous ``right extension'' 
$(\xi_M/\xi_L) \colon L \to M$ 
of $\xi_M$ along $\xi_L$,
\textit{i.e.},
the map $(\xi_M/\xi_L)$ satisfies
$(\xi_M/\xi_L) \circ \xi_L \leq \xi_M$,
and if 
$h \colon L \to M$ is a continuous map which satisfies 
$h \circ \xi_L \leq \xi_M$
then 
$h \leq (\xi_M/\xi_L)$.
If any map witnesses the tightening relation, then so does $(\xi_M/\xi_L)$.
But the map $R_{L,M} \colon M^Y \to M^L$ which sends $h \in M^Y$ to $(h/\xi_L) \in M^L$ is discontinuous in general.
We can therefore in general not expect $\widetilde{\nu_Z \circ \varphi}$ 
in Proposition \ref{Proposition: extending global probe, non-uniform} 
to depend continuously on $i$ and $\varphi$.
Computability of $\widetilde{i \circ \varphi}$ may fail even point-wise.
The example \cite[Example 4.32]{NeumannPhD} shows that there exist computable $i$ and $\varphi$
such that all continuous functions $\widetilde{i \circ \varphi}$
as in Proposition \ref{Proposition: extending global probe, non-uniform}
are uncomputable.

On the other hand, if $L$ is such that $R_{L,M}$ is continuous for all effective injective spaces $M$
then we do obtain effective versions of Propositions \ref{Proposition: extending global probe, non-uniform} and \ref{Proposition: extending probe, non-uniform}.
Escard\'o \cite{EscardoInjective} has shown that $R_{L,M}$ is continuous if and only if the upper adjoint of 
$\xi_L^* \colon \O(L) \to \O(Y)$ is continuous, \textit{i.e.}, if $\xi_L$ is a proper map in the sense of Hofmann and Lawson \cite{HofmannLawson}.
This constitutes arguably too much of a restriction.
For instance, it precludes the choice of the natural approximation space 
$\nu_X \colon X \to \O^2(X)$
for most spaces $X$.
See \cite[pp. 59 -- 66]{NeumannPhD} for further discussion.

However, for the purpose of obtaining an effective version of Propositions \ref{Proposition: extending global probe, non-uniform} and \ref{Proposition: extending probe, non-uniform} it is not necessary to compute the greatest continuous extension in the above sense, since a smaller extension may already witness the relevant tightening relation.
Let us attempt to capture this with a very general definition:
\begin{defi}\label{Definition: uniformly envelopable}
    Let $f \colon X \to Y$ be a function between effective $T_0$ spaces.
    We call $f$ \emph{uniformly envelopable} if
    there exists an envelope 
    $F \colon X \to L$ 
    of 
    $f \colon X \to Y$
    such that for all effective injective spaces $M$ there exists a continuous map 
    \[
        E_{L,M} \colon M^Y \to M^L
    \]
    satisfying 
    $E_{L,M}(\varphi) \circ \xi_L \leq \varphi$
    for all $\varphi \colon Y \to M$,
    and for all envelopes 
    $G \colon X \to M$
    of $f$ we have 
    $E_{L,M}(\xi_M) \circ F \geq G$.
\end{defi}

As mentioned above, if $f$ has an envelope 
$F \colon X \to L$ 
such that 
$\xi_L^* \colon \O(L) \to \O(Y)$
has a continuous upper adjoint,
then $f$ is uniformly envelopable and we may put 
$E_{L,M}(\varphi) = \varphi/\xi_L$.

But the existence of a continuous upper adjoint is not necessary for uniform envelopability.
To any function $f \colon X \to Y$ we can assign the principal $\O^2(Y)$-envelope 
$F \colon X \to \O^2(Y)$.
Given a function $\varphi \colon Y \to M$ with values in an effective injective space $M$
we can (relatively) compute the function 
$\rho_M \circ \varphi^{**} \colon \O^2(Y) \to M$.
This yields the function 
\begin{equation}\label{eq: O^2 extension}
    E_{\O^2(Y), M} \colon M^Y \to M^{\O^2(Y)},
    \;
    \varphi \mapsto \rho_M \circ \varphi^{**}
\end{equation}
with 
$E_{\O^2(Y), M}(\varphi) \circ \nu_Y = \varphi$
for all 
$\varphi \colon Y \to M$.
In general the extension 
$E_{\O^2(Y), M}(\varphi)$
is not equal to the greatest continuous extension of $\varphi$ along $\nu_Y$.
Yet, it is still possible for $F$ to tighten all envelopes $G \colon X \to M$ of $f$ via $E_{\O^2(Y),M}(\xi_M)$.

It turns out that this special case already characterises uniform envelopability.
We recall a central definition:
A robust property of $f \colon X \to Y$ at $x \in X$, or -- by abuse of notation -- a robust property of $f(x)$ is an open set $V \in \O(Y)$ such that $f^{-1}(V)$ is a neighbourhood of $x$.

\begin{thm}\label{Theorem: motivation}
    Let $f \colon X \to Y$ be a function between effective $T_0$ spaces.
    Let $F \colon X \to \O^2(Y)$ be its principal $\O^2(Y)$-envelope.
    Then the following are equivalent:
    \begin{enumerate}
        \item $f$ is uniformly envelopable.
        \item For all robust properties $V$ of $f(x)$ we have $V \in F(x)$.
        \item 
        $F$
        tightens all envelopes 
        $G \colon X \to M$ 
        of $f$ via 
        $E_{\O^2(Y),M}(\xi_M)$,
        where 
        $E_{\O^2(Y),M}$
        is defined as in \eqref{eq: O^2 extension}.
    \end{enumerate}
\end{thm}
\begin{proof}
    Assume that $f$ is uniformly envelopable.
    Let 
    $H \colon X \to L$ 
    be an envelope of $f$ such that there exist maps 
    $E_{L,M} \colon M^Y \to M^L$ 
    with
    $E_{L, M}(\varphi) \circ \xi_L \leq \varphi$
    such that 
    $H$ 
    tightens all envelopes 
    $G \colon X \to M$ 
    of 
    $f$
    via 
    $E_{L,M}(\xi_M)$.
    Consider the map 
    $E_{L,\Sigma} \colon \Sigma^{Y} \to \Sigma^L$.
    Write this map as 
    $s \colon \O(Y) \to \O(L)$.
    Write
    $\chi_U \colon Y \to \Sigma$
    for the characteristic function 
    of the open set $U \in \O(Y)$.
    We have by definition
    $
        E_{L,\Sigma}(\chi_{U}) \circ \xi_L
        \leq \chi_{U},
    $
    for all $U \in \O(Y)$,
    which translates to 
    $
        \xi_L^* \circ s \leq \id_{\O(Y)}
    $.

    Consider the map 
    $s^* \colon \O^2(L) \to \O^2(Y)$.
    Then 
    $s^* \circ \nu_L \circ H$
    is an $\O^2(Y)$-envelope of $f$, for 
    \[
        s^* \circ \nu_L \circ H 
        \leq s^* \circ \nu_L \circ \xi_L \circ f 
        = s^* \circ \xi_L^{**} \circ \nu_Y \circ f 
        \leq \nu_Y \circ f.
    \] 
    It follows that $s^* \circ \nu_L \circ H \leq F$.
    
    Let $U \in \O(Y)$ be a robust property of $f(x)$.
    We can view 
    $\chi_U \colon Y \to \Sigma$
    as an approximation space over $Y$.
    Then there exists an envelope 
    $G \colon X \to \Sigma$ 
    with values in this approximation space
    such that 
    $G(x) = \top$.
    It follows that $H(x) \in s(U)$.
    Hence 
    $U \in s^* \circ \nu_L \circ H(x)$.
    Thus a fortiori $U \in F(x)$.

    Now assume that $V \in F(x)$ for all robust properties $U$ of $f(x)$.
    Let $G \colon X \to M$ be an envelope of $f$.
    We claim that $F$ tightens $G$ via 
    $E_{\O^2(Y),M}(\xi_M) = \rho_M \circ \xi_M^{**}$.
    Observe that we have 
    \[
        \rho_M \circ \xi_M^{**} \circ \nu_Y = \rho_M \circ \nu_M \circ \xi_M = \xi_M.
    \]
    Thus, it remains to verify the inequality
    $\rho_M \circ \xi_M^{**} \circ F \geq G$.
    Let $G(x) \in W$.
    Then, since $G$ is an envelope, $\xi_M^{*}(W)$ is a robust property of $f(x)$.
    By assumption we obtain $\xi_M^{*}(W) \in F(x)$.
    Hence $W \in \xi_M^{**} \circ F(x)$.
    It follows that 
    $\xi_M^{**} \circ F \geq \nu_M \circ G$
    and hence 
    $\rho_M \circ \xi_M^{**} \circ F \geq G$.

    Finally, if $F$ tightens all envelopes $G$ via $E_{\O^2(Y),M}(\xi_M)$
    then $f$ is uniformly envelopable by definition.
\end{proof}

Theorem \ref{Theorem: motivation} is nothing more than a simple observation. 
Yet it has remarkable consequences:
firstly, it reduces the task of finding an envelope 
$F \colon X \to L$ 
and maps 
$E_{L,M}$
as in Definition \ref{Definition: uniformly envelopable}
to the rather more tractable task of checking a fairly concrete property of the
principal $\O^2(Y)$-envelope.
Secondly, we make hardly any assumptions on the maps $E_{L,M}$ in Definition \ref{Definition: uniformly envelopable}.
By contrast, $\O^2$ is a monad on the category of effective $T_0$ spaces and the extensions $E_{\O^2(Y),M}$ are all obtained in the same way,
by applying the functor $\O^2$ and then the retraction $\rho_M$.
We thus obtain a lot of extra structure, which deserves further investigation.


In this paper, we introduce \emph{uniform envelopes}, which constitute mild generalisations of principal $\O^2$-envelopes.
A uniform envelope $F \colon X \to L$ of $f \colon X \to Y$ is equipped with a map $u_L \colon L \to \O^2(Y)$ that satisfies certain axioms which connect it to the monad structure of $\O^2$.
As above, this yields a family of maps $E_{L,M} \colon M^Y \to M^L$ for all injective spaces $M$.
The uniform envelope $F$ tightens an (arbitrary) envelope $G \colon X \to M$ of $f$ if it tightens $G$ via the map $E_{L,M}(\xi_M)$.
The uniform envelope $F$ is uniformly universal if it uniformly tightens all envelopes $G$ of $f$.
It follows essentially immediately from the definition that if 
$F$ is uniformly universal 
and 
$\varphi \colon Y \to Z$ 
is any continuous map then 
$E_{L,\O^2(Z)}(\nu_Z \circ \varphi) \circ F$
is a uniformly universal envelope of $\varphi \circ f$.

More generally, a uniform envelope $F \colon X \to L$ of $f \colon X \to Y$ 
and a uniform envelope $G \colon Y \to M$ of $g \colon Y \to Z$ can be composed
to yield an envelope $G \bullet F$ of $g \circ f$.
If $g$ is not continuous, then $G \bullet F$ may fail to be uniformly universal.
We can explain this failure and identify situations in which it does not occur by studying how open sets of the 
co-domain can be transported back along a function and its universal envelope.
An arbitrary function $f \colon X \to Y$ transports open sets of $Y$ to open sets of $X$ via the -- not necessarily continuous -- function
\[ 
	\intr{f} \colon \O(Y) \to \O(X),
	\; 
	\intr{f}(V) = \intr{f^{-1}(V)}.
\]
A continuous function $F \colon X \to \O^2(Y)$ transports open sets of $Y$ to open sets of $X$ via the continuous function
\[
	\starred{F} \colon \O(Y) \to \O(X),
	\; 
	\starred{F}(V) = 
		\Set{x \in X}{F(x) \ni V}.
\]
We show that $F$ is uniformly universal if and only if $\starred{F} = \intr{f}$.
We further show that $\starred{(G \bullet F)} = \starred{F} \circ \starred{G}$
for all continuous maps $F \colon X \to \O^2(Y)$ and $G \colon Y \to \O^2(Z)$.
Thus, $G \bullet F$ is a uniformly universal envelope of $g \circ f$ if and only if $\intr{(g \circ f)} = \intr{f} \circ \intr{g}$.
A sufficient condition for the latter is that $f$ is open or that $g$ is continuous.
We establish a partial converse to this.
We also show that in general the problem of proving that  
$\intr{(g \circ f)} = \intr{f} \circ \intr{g}$
can be extremely hard by reducing the hyperbolicity conjecture from complex dynamics to a specific instance of this problem.

It is natural to seek for a-priori criteria that guarantee uniform envelopability. 
We show that a space $Y$ has the property that every function $f \colon X \to Y$ is uniformly envelopable if and only if it is Noetherian.
Dually, we show that a space $X$ has the property that every function $f \colon X \to Y$ is uniformly envelopable if and only if $X$ contains no infinite compact sets\footnote{An earlier version of this paper contained a weaker version of this latter claim. The present version of the result was obtained thanks to a suggestion by an anonymous referee.}.

In practice, a common way of relaxing the problem of computing a function 
$f \colon X \to Y$,
where $X$ is a metric space, 
is to consider the ``approximate'' problem
\[
    \widetilde{f} \colon X \times \Q_+ \rightrightarrows Y,
	\;
	\widetilde{f}(x,\varepsilon) = 
		\Set{y \in Y}{\exists \tilde{x} \in B(x,\varepsilon). y = f(\tilde{x})}
\]
where $\Q_+$ denotes the space of positive rational numbers.
This relaxation underlies for instance 
the non-deterministic equality tests used in exact real computation,
backwards-stable algorithms in numerical analysis, 
and the definition of approximate solutions to non-linear equations.
If $\alpha \colon \widetilde{X} \to X$ 
and $\beta \colon \widetilde{X} \times Y \to Z$
are continuous functions where $Z$ is a metric space
then a relativised algorithm for $\widetilde{f}$ can be made into a relativised algorithm for $f$
with the help of a relatively computable multi-valued \cite[Section 1.4, pp. 10--11]{WeihrauchBook} map 
$
    \omega \colon \widetilde{X} \times \Q_{+} \rightrightarrows \Q_{+}
$
satisfying
\[
	\forall \widetilde{x} \in \widetilde{X}.
	\forall \varepsilon \in \Q_+.
	\forall x \in X.
	\forall \delta \in \omega(\widetilde{x}, \varepsilon).
	\left(
		d(x, \alpha(\widetilde{x})) < \delta 
		\rightarrow
		d(\beta(\widetilde{x}, f \circ \alpha (\widetilde{x})),
		 \beta(\widetilde{x}, f(x))) < \varepsilon
	\right).
\]
An analogous result to 
Proposition \ref{Proposition: extending probe, non-uniform} 
shows that if $\widetilde{X}$ is a complete metric space, then 
such a function $\omega$ exists if and only if $\beta$ 
extends as in 
Proposition \ref{Proposition: extending probe, non-uniform}
to 
$\widetilde{X} \times L$
up to embedding the co-domain $Z$ into $\K([0,1]^{\N})$.
But in order for this result to hold true we do not require the envelope $F$ to be universal for the class of all envelopes.
We introduce the class of regular approximation spaces, which have separation properties analogous to those of regular Hausdorff spaces.
An envelope is $\mathcal{R}$-universal if it tightens all envelopes with values in a regular approximation space.
A uniform envelope is uniformly $\mathcal{R}$-universal if it uniformly tightens all such envelopes.
The aforementioned analogue to Proposition \ref{Proposition: extending probe, non-uniform} already holds true for $\mathcal{R}$-universal envelopes.
One also easily obtains an analogue to 
Proposition \ref{Proposition: extending global probe, non-uniform}
involving a similar modulus.
The effective version of these results involving uniformly $\mathcal{R}$-universal envelopes yields a somewhat effective criterion for the existence of a modulus $\omega$ as above:
Given 
$\alpha \colon \widetilde{X} \to X$ 
and 
$\beta \colon \widetilde{X} \times Y \to Z$, 
fix an embedding $i \colon Z \to [0,1]^{\N}$,
compute the extension $\widetilde{i \circ \beta} \colon \widetilde{X} \times L \to \K([0,1]^{\N})$,
and check if the functions
$\widetilde{i \circ \beta}(\widetilde{x}, F \circ \alpha (\widetilde{x}))$
and 
$i \circ \beta(\widetilde{x}, f \circ \alpha(\widetilde{x}))$
are the same.

The notion of $\mathcal{R}$-universality is a useful weakening of that of universality since, unlike in the case of uniform envelopability, we have a simple and powerful sufficient criterion available that asserts uniform $\mathcal{R}$-envelopability.
If $f \colon X \to Y$ is a function with values in a countably based effective Hausdorff space such that there exists a continuous map 
$B \colon X \to \K(Y)$ satisfying $f(x) \in B(x)$ for all $x \in X$
then the principal $\O^2(Y)$-envelope is uniformly $\mathcal{R}$-universal.

Uniform $\mathcal{R}$-universality has a similar characterisation as uniform universality in terms of pullback operations on continuous maps.
Let $\mathcal{C}_X(-)$ denote the sheaf of continuous real-valued functions on a space $X$.
Then any function $f \colon X \to Y$ defines a function 
\[
    \intr{f}_{\R} \colon \Sheaf{Y}{-} \to \Sheaf{X}{-}
\] 
which sends a section 
$\varphi \colon V \to \R$ 
to 
$\varphi \circ f \colon U \to \R$
where $U$ is the largest open set such that $\varphi \circ f$ is well-defined and continuous.
Similarly, a continuous map 
$F \colon X \to \O^2(Y)$
defines a function 
\[
    \starred{F}_{\R} \colon \Sheaf{Y}{-} \to \Sheaf{X}{-}.
\] 
We show that if $Y$ is a normal Hausdorff space then $F$ is uniformly $\mathcal{R}$-universal if and only if 
$\left(\starred{F}_{\R}\right)|_{\mathcal{C}_Y(Y)} = \left(\intr{f}_{\R}\right)|_{\mathcal{C}_Y(Y)}$,
where $\mathcal{C}_{Y}(Y)$ denotes the set of global sections.
This characterisation at the same time reveals that uniformly $\mathcal{R}$-universal envelopes have worse composition behaviour than 
uniformly universal envelopes, since $\starred{F}$ need not transport global sections to global sections. 
We show that the composition
$G \bullet F$
of two $\mathcal{R}$-universal envelopes is again uniformly $\mathcal{R}$-universal if and only if the functions
$\intr{(g \circ f)}_{\R}$
and 
$\intr{f}_{\R} \circ \intr{g}_{\R}$
agree on global sections.
We obtain the following sufficient criterion:
If $F$ is a uniformly universal envelope of $f$ 
and $G$ is a uniformly $\mathcal{R}$-universal envelope of $g$ 
and additionally $f$ is an open map then $G \bullet F$ is again uniformly $\mathcal{R}$-universal.
We show that the assumption that $F$ be uniformly universal cannot be weakened in general.

Finally, we observe that we can assign to any function $f \colon X \to Y$ a canonical envelope 
that generalises the principal $\O^2(Y)$-envelope of a uniformly envelopable function.
On such canonical envelopes we can define a composition operation that generalises the composition of uniform envelopes.

A function 
$f \colon X \to Y$
is uniformly envelopable if and only if the induced function 
$\intr{f} \colon \O(Y) \to \O(X)$
is continuous.
This says that given a point $x \in X$ and a robust property $V$ of $f(x)$
we can verify that $f(x) \in V$.

If the function $\intr{f}$ is discontinuous, then a name of 
a set $V \in \O(Y)$ does not provide enough information to 
verify that $f(x) \in V$ whenever $V$ is a robust property of 
$f(x)$.
Extra information on $V$ is needed to perform this verification.
In general the same open set $V$ can be a robust property of 
multiple points $x \in X$, 
and the amount of extra information required to verify 
that $f(x) \in V$ may depend on $x$.

To keep track of this we propose the following definition:
An \emph{advice bundle} for $f$ is an effective injective space 
$A$ together with a continuous map
$\pi_A \colon A \to \O(Y)$
which preserves arbitrary joins and finite meets 
and has a continuous section.
For any advice bundle $\pi_A \colon A \to \O(Y)$
there exists a greatest continuous map 
$\starred{F} \colon A \to \O(X)$
satisfying
$\starred{F} \leq \intr{f} \circ \pi_A$.
We call maps of this kind \emph{co-envelopes}.
Co-envelopes
$\starred{F} \colon A \to \O(X)$
and 
$\starred{G} \colon B \to \O(X)$ 
of the same function can be compared as follows:
$\starred{F}$
tightens 
$\starred{G}$
if there exists a continuous map 
$\Phi \colon B \to A$
satisfying 
$\pi_A \circ \Phi = \pi_B$
and 
$\starred{F} \circ \Phi \geq \starred{G}$.
The induced pre-order is isomorphic to the pre-order of envelopes of $f$.
In particular every function has a universal co-envelope.

An advice bundle $\pi_E \colon E \to \O(Y)$ with principal co-envelope 
$\starred{F} \colon E \to \O(X)$
is a \emph{least advice bundle} for $f$ if 
for all advice bundles 
$\pi_A \colon A \to \O(Y)$ 
such that the associated principal co-envelope 
$\starred{G} \colon A \to \O(X)$
is universal,
there exists a unique continuous map 
$r \colon A \to E$
which admits a continuous section 
$s \colon E \to A$,
such that 
$r$ 
witnesses the tightening of 
$\starred{G}$
by 
$\starred{F}$
and $s$ witnesses the tightening in the other direction.

We show that every function $f$ has a least advice bundle $\mathfrak{A}_f$.
Since being a least advice bundle is a universal property, this advice bundle is unique up to unique isomorphism.
It is isomorphic to $\O(Y)$ if and only if $f$ is uniformly envelopable.

This allows us to assign to every function $f$ a canonical co-envelope 
$\starred{\mathfrak{E}_f} \colon \mathfrak{A}_f \to \O(X)$
which we call the \emph{primary co-envelope}.
It induces an envelope 
$\mathfrak{E}_f \colon X \to \O(\mathfrak{A}_f)$ of $f$,
called the \emph{primary envelope}.

In a sense, the least advice bundle makes the task of finding the greatest continuous lift of a function of type 
$\varphi \colon \O(Z) \to \O(Y)$
along $\pi_{\mathfrak{A}_f} \colon \mathfrak{A}_f \to \O(Y)$
``as easy as possible''
by ensuring that $\varphi$ has as few continuous lifts as possible.
Thus, while we do not obtain effective versions of 
Propositions 
\ref{Proposition: extending global probe, non-uniform} 
and 
\ref{Proposition: extending probe, non-uniform} 
in general,
we obtain versions of these results that are ``as close to effective''
as one can expect them to be.

Primary co-envelopes can be composed analogously to uniform envelopes.
We obtain a composition theorem that is similar to the previous ones:
If $g$ is continuous or $f$ is open then the composition 
$\starred{F} \bullet \starred{G}$ 
of the primary co-envelopes of $g$ and $f$
is the principal $\mathfrak{A}_g$-co-envelope 
of $g \circ f$.
This conclusion is much weaker than that of the previous results.
We show that it cannot be improved in general.

\section{Uniform Envelopes}

Recall that $\O^2$ is a monad on the category of effective $T_0$ spaces with 
unit $\nu_Y \colon Y \to \O^2(Y)$ and multiplication 
$\nu_{\O(Y)}^* \colon \O^4(Y) \to \O^2(Y)$.
We also write $\mu_Y \colon \O^4(Y) \to \O^2(Y)$ for the multiplication.

            A \emph{uniform approximation space} for $Y$ is an approximation space 
            $\xi_L \colon Y \to L$ 
            together with a map 
            $u_L \colon L \to \O^2(Y)$
            satisfying the following three axioms:
            \begin{enumerate}
                \item $u_L \circ \xi_L \leq \nu_Y$
                \item $\rho_L \circ \xi_L^{**} \circ u_L \geq \id_L$
                \item $u_L \circ \rho_L = \mu_{Y} \circ u_L^{**}$.
            \end{enumerate}
            This defines for each effective injective space $M$ a map 
            \[
                E_{L,M} \colon M^Y \to M^L,
                \; 
                \varphi \mapsto \rho_M \circ \varphi^{**} \circ u_L.
            \]
            Let $f \colon X \to Y$.
            A \emph{uniform envelope of $f$}
            is an envelope $F \colon X \to L$ 
            with values in a uniform approximation space.
            Let $F\colon X \to L$ be a uniform envelope of $f$ 
            and let $G \colon X \to M$ be an envelope of $f$.
            Then $F$ \emph{uniformly tightens} $G$ if it tightens $G$ via the map $E_{L,M}(\xi_M)$.
            We call $F$ and $G$ \emph{uniformly equivalent} if they uniformly tighten each other.
        Let $F\colon X \to L$ be a uniform envelope of $f$.
        Then $F$ is called \emph{uniformly universal} if it uniformly tightens all
        envelopes of $f$.

        More generally, if $\mathcal{C}$ is a class of approximation spaces then 
        $F$ is called \emph{uniformly $\mathcal{C}$-universal}
        if it uniformly tightens all envelopes of $f$
        with values in a space that belongs to $\mathcal{C}$.

Clearly, $\O^2(Y)$ with uniformity map $\id_{\O^2(Y)}$ is a uniform approximation space for every effective $T_0$ space $Y$.
One readily verifies that $E_{L,M}(h) \circ \xi_L \leq h$ for all functions $h \colon Y \to M$.
In particular, if $F$ uniformly tightens $G$ then $F$ tightens $G$.
It follows immediately from the definition that the uniform tightening relation is reflexive and transitive.

Uniform envelopes generalise the principal $\O^2(Y)$-envelope.
In terms of information content this generalisation is not substantial: 

\begin{prop}\label{Proposition: uniform envelope equivalent to O^2-envelope}
    Let $f \colon X \to Y$ be a function between effective $T_0$ spaces.
    Let $F \colon X \to L$ be a uniform envelope of $f$.
	Then $F$ is uniformly equivalent to an $\O^2(Y)$-envelope.
	In particular $F$ is uniformly tightened by the principal $\O^2(Y)$-envelope.
\end{prop}
\begin{proof}
    Consider the map 
    $u_L \circ F \colon X \to \O^2(Y)$.
    This map is an $\O^2(Y)$-envelope of $f$, for 
    \[ 
        \nu_Y \circ f 
        \geq u_L \circ \xi_L \circ f 
        \geq u_L \circ F.
    \]
	We have $\nu_Y^* \circ \nu_{\O(Y)} = \id_{\O(Y)}$
    and hence 
	$
		E_{L,\O^2(Y)}(\nu_Y)
		= 
		\nu_{\O(Y)}^* \circ \nu_Y^{**} \circ u_L 
		= u_L
	$.
    It follows that $F$ uniformly tightens this envelope.
	Conversely,
	\[
	E_{\O^2(Y),L}(\xi_L) \circ u_L \circ F
	=
	\rho_L \circ \xi_L^{**} \circ u_L \circ F
	\geq 
	F.
	\]
	Hence $F$ and $u_L \circ F$ are uniformly equivalent.
	The addendum follows immediately.
\end{proof}

While uniform envelopes cannot contain more information than the principal $\O^2(Y)$-envelope,
they can be more convenient to work with.
For instance, one is often interested in studying the lower semi-continuity of set-valued maps 
$f \colon X \rightrightarrows Y$.
This can be modelled by studying the continuity of the map 
$f \colon X \to \V(Y)$.
The space $\O^2(\V(Y))$ is reasonably complicated already.
On the other hand, $\V(Y)$ embeds into $\O^2(Y)$.
This turns out to be a uniform approximation space:

\begin{prop}
    Let $Y$ be an effectively countably based effective $T_0$ space.
    Then $\O^2(Y)$ with inclusion map 
    \[
        i \colon \V(Y) \to \O^2(Y),
        \; 
        i(A) = \Set{U \in \O(Y)}{U \cap A \neq \emptyset}
    \]
    and uniformity map 
    \[
        j \colon \O^2(Y) \to \O^2(V(Y)),
        \; 
        j(\mathcal{U}) 
        = \Set{V \in \O(\V(Y))}
              {V \supseteq \Set{A \in \V(Y)}{\forall U \in \mathcal{U}. (A\cap U \neq \emptyset)}}
    \] 
    is a uniform approximation space for $\V(Y)$.

    Let $f \colon X \to \V(Y)$.
    Then the principal $\O^2(\V(Y))$-envelope of $f$ is uniformly universal if and only if the principal 
    $\O^2(Y)$-envelope is.
\end{prop}
\begin{proof}
    It is shown in \cite[Proposition 2.56]{NeumannPhD} that the map 
    \[
        j \colon 
        \O^2(Y) \to \K(\V(Y))
        \;
        \mathcal{U}
        \mapsto
        \Set{A \in \V(Y)}{\forall U \in \mathcal{U}. (A \cap U \neq \emptyset)}
    \]
    is well-defined and continuous.
    An easy calculation shows that 
    $j \circ i(A) = \nu_{\V(Y)}(A)$.
    Similarly, a straightforward but cumbersome calculation shows 
    $\mu_Y \circ i^{**} \circ j \geq \id$.
    The third axiom 
    $j \circ \mu_Y = \mu_{\V(Y)} \circ j^{**}$
    is just the naturality of $\mu$. 

    Uniform universality of the principal $\O^2(Y)$-envelope implies uniform universality of the principal $\O^2(\V(Y))$-envelope 
    by Proposition \ref{Proposition: uniform envelope equivalent to O^2-envelope}.
    The converse direction is given in \cite[Lemma 4.47]{NeumannPhD}.
\end{proof}

\begin{defi}
	Let $X$, $Y$, and $Z$ be computable $T_0$ spaces.
	Let 
    $F\colon X \to L$ 
    and 
    $G \colon Y \to M$ 
    be envelopes with inclusion maps 
	$\xi_L \colon Y \to L$ 
    and 
    $\xi_M \colon Z \to M$.
	Assume that $F$ is a uniform envelope.
	The \emph{composition} 
    $G \bullet F$ 
    is defined as 
	$G \bullet F(x) = E_{L,M}(G) \circ F$.
\end{defi}

It is straightforward to verify that the composition of two uniform envelopes $G \bullet F$ is a uniform envelope of the composition $g \circ f$.
We will show that the composition respects the uniform tightening ordering.
We need a technical result as a preparation which is easily established:

\begin{lem}\label{Lemma: characterisation of uniform tightening between uniform envelopes}
	Let $F \colon X \to L$ and $G\colon X \to M$ be uniform envelopes
    of the same function $f \colon X \to Y$.
	Then $F$ uniformly tightens $G$ if and only if 
	$
		u_L \circ F \geq u_M \circ G.
	$
\end{lem}

\begin{prop}\label{Proposition: composition respects order}
	Let $X$, $Y$, and $Z$ be computable $T_0$ spaces.
	For $i = 1,2$ let
	$F_i\colon X \to L_i$ 
	and 
	$G_i \colon Y \to M_i$ 
	be uniform envelopes with inclusion maps 
	$\xi_L^i \colon Y \to L_i$ and $\xi_M^i \colon Z \to M_i$.
	If $F_1$ uniformly tightens $F_2$
	and $G_1$ uniformly tightens $G_2$
	then 
	$G_1 \bullet F_1$
	uniformly tightens
	$G_2 \bullet F_2$. 
\end{prop}
\begin{proof}
	Our aim is to prove that 
	$E_{M_1,M_2}(\xi_{M_2}) \circ E_{L_1,M_1}(G_1) \circ F_1 \geq E_{L_2,M_2}(G_2) \circ F_2$.
	Expand the left hand side according to the definition:
	\[
		E_{M_1,M_2}(\xi_{M_2}) \circ E_{L_1,M_1}(G_1) \circ F_1
		= 
		\rho_{M_2} \circ \xi_{M_2}^{**} \circ u_{M_1} \circ \rho_{M_1} \circ G_1^{**} \circ u_{L_1} \circ F_1.
	\]
	Use $u_{M_1} \circ \rho_{M_1} = \mu_Z \circ u_{M_1}^{**}$:
	\[
		E_{M_1,M_2}(\xi_{M_2}) \circ E_{L_1,M_1}(G_1) \circ F_1
		=
		\rho_{M_2} \circ \xi_{M_2}^{**} \circ \mu_Z \circ u_{M_1}^{**} \circ G_1^{**} \circ u_{L_1} \circ F_1.
	\]
	By Lemma \ref{Lemma: characterisation of uniform tightening between uniform envelopes} we have 
	$u_{M_1} \circ G_1 \geq u_{M_2} \circ G_2$
	and 
	$u_{L_1} \circ F_1 \geq u_{L_2} \circ F_2$
	and hence:
	\[
		E_{M_1,M_2}(\xi_{M_2}) \circ E_{L_1,M_1}(G_1) \circ F_1
		\geq
		\rho_{M_2} \circ \xi_{M_2}^{**} \circ \mu_Z \circ u_{M_2}^{**} \circ G_2^{**} \circ u_{L_2} \circ F_2.
	\]
	Translate back using $\mu_Z \circ u_{M_2}^{**} = u_{M_2} \circ \rho_{M_2}$:
	\begin{align*}
		E_{M_1,M_2}(\xi_{M_2}) \circ E_{L_1,M_1}(G_1) \circ F_1
		&\geq
		\rho_{M_2} \circ \xi_{M_2}^{**} \circ u_{M_2} \circ \rho_{M_2} \circ G_2^{**} \circ u_{L_2} \circ F_2 \\ 
		&\geq \rho_{M_2} \circ G_2^{**} \circ u_{L_2} \circ F_2\\ 
		&= E_{L_2,M_2}(G_2) \circ F_2.
	\end{align*}
	Thus the claim is shown.
\end{proof}

Finally we observe that composition is associative up to equivalence with respect to the uniform tightening relation.

\begin{prop}
	Let
	$F\colon X \to L$,
	$G \colon Y \to M$,
	and 
	$H \colon Z \to N$ 
	be uniform envelopes of 
    $f \colon X \to Y$,
    $g \colon Y \to Z$,
	and 
	$h \colon Z \to W$
	respectively.
	Then $H \bullet (G \bullet F)$ 
    and $(H \bullet G) \bullet F$
    are uniformly equivalent.
\end{prop}
\begin{proof}
	By Proposition \ref{Proposition: uniform envelope equivalent to O^2-envelope} every uniform envelope of $f \colon X \to Y$ is uniformly equivalent to an envelope with values in the approximation space $\O^2(Y)$.
	By Proposition \ref{Proposition: composition respects order} this equivalence is preserved by composition.
	We may thus assume up to equivalence that $F$, $G$, and $H$ take values in the approximation spaces $\O^2(Y)$, $\O^2(Z)$, and $\O^2(W)$ respectively.
	But then the composition is the same as the composition in the Kleisli category of the monad $\O^2$, which is of course associative.
\end{proof}

When working with functions whose co-domains are effective Hausdorff\footnote{If $Y$ is not Hausdorff, then $\Kb(Y)$ may fail to admit binary joins, so that it may fail to be a lattice. See e.g.~\cite[p. 48]{NeumannPhD} for an example.} spaces, it is often more convenient and intuitive to work in the lattice $\Kb(Y)$ rather than $\O^2(Y)$.
The following results allow one to do so:

\begin{prop}
    Let $f \colon X \to Y$ be a uniformly envelopable function between effective $T_0$ spaces.
    Assume that $Y$ is an effective Hausdorff space.
    Let 
    $F \colon X \to \O^2(Y)$
    be the principal $\O^2(Y)$-envelope of $f$.
    Let 
    $G \colon X \to \Kb(Y)$
    be the greatest continuous approximation of $f$ with values in $\Kb(Y)$.
    Then 
    \[
        G(x) = \bigcap_{U \in F(x)} U
    \]
    and 
    \[
        F(x) = \Set{U \in \O(Y)}{Y \supseteq G(x)}.
    \]
\end{prop}
\begin{proof}
    Since $f$ is assumed to be uniformly envelopable, the envelope 
    $F \colon X \to \O^2(Y)$ 
    witnesses all robust properties of $f$
    in the sense of Theorem \ref{Theorem: motivation}.
    Since the robust properties of $f(x)$ form a filter, it follows that 
    $F(x)$ is a filter for all $x \in X$.
    The result now follows from the Hofmann-Mislove theorem \cite[Theorem II-1.20]{ContinuousLattices}.
\end{proof}

\begin{prop}
    Let $X$, $Y$, and $Z$ be effective $T_0$ spaces.
    Let 
    $F \colon X \to \Kb(Y)$, 
    $G \colon Y \to \Kb(Z)$.
    Let 
    \[
        i_Y \colon \Kb(Y) \to \O^2(Y),
        \;
        i_Y(K) = \Set{U \in \O(Y)}{U \supseteq K}.
    \]
    Then the composition
    $(i_Z \circ G) \bullet (i_Y \circ F)$
    is equal
    to 
    $i_Z \circ (G \circ F)$,
    where
    $G \circ F$ 
    denotes the composition in the Kleisli category of the monad $\Kb$.
\end{prop}

We will now study the question when the composition of two uniformly universal envelopes is a uniformly universal envelope of the composition.
In view of Propositions \ref{Proposition: uniform envelope equivalent to O^2-envelope} and \ref{Proposition: composition respects order} we may restrict our discussion to principal $\O^2$-envelopes.

Let $\xi_L \colon Z \to L$ be an approximation space for $Z$.
Let $g \colon Y \to Z$ be a continuous map.
The \emph{pullback} $g^*L$ of $L$ along $g$ is the approximation space $\xi_L \circ g \colon Y \to L$ for $Y$.
If $F \colon X \to L$ is an envelope of $g\circ f \colon X \to Z$ 
then this yields a pullback envelope 
$g^*F\colon X \to g^*L$
of 
$f \colon X \to Y$.

We say that a class of approximation spaces $\mathcal{C}$ is closed under pullbacks along $g \colon Y \to Z$ 
if for all approximation spaces 
$\xi_L \colon Z \to L$ 
the pullback $g^* L$ belongs to $\mathcal{C}$.

\begin{prop}\label{Proposition: post-composing continuous maps}
	Let $f \colon X \to Y$ be a function between computable $T_0$ spaces.
	Let $\mathcal{C}$ be a class of approximation spaces.
	Let $g \colon Y \to Z$ be a continuous map such that $\mathcal{C}$ is closed under pullbacks along $g$.
	Let $F \colon X \to L$ be a uniformly $\mathcal{C}$-universal envelope.
	Assume that 
    $\nu_Z \colon Z \to \O^2(Z)$ 
    is an approximation space in 
    $\mathcal{C}$.
	Then
	$g^{**} \circ u_L \circ F \colon X \to \O^2(Z)$
	is a uniformly $\mathcal{C}$-universal envelope of $g \circ f$.
	Here, as usual, $\O^2(Z)$ is made into a uniform approximation space by letting $u_{\O^2(Z)} = \id_{\O^2(Z)}$.
\end{prop}
\begin{proof}
	Let $G \colon X \to M$ be a $\mathcal{C}$-envelope of $g \circ f$.
	Then by assumption on $g$, the envelope $g^* G \colon X \to g^* M$ is a $\mathcal{C}$-envelope of $f$.
	By uniform universality of $F$ we have $E_{L,M}(\xi_M \circ g) \circ F \geq G$.
	Now, 
	\begin{align*}
		E_{L,M}(\xi_M \circ g) 
		&= \rho_M \circ (\xi_M \circ g)^{**} \circ u_L\\ 
		&= \rho_M \circ \xi_M^{**} \circ g^{**} \circ u_L\\
        &= \rho_M \circ \xi_M^{**} \circ u_{\O^2(Z)} \circ g^{**} \circ u_L\\
        &= E_{\O^2(Z), M}(\xi_M) \circ g^{**} \circ u_L.
    \end{align*}
    It follows that $g^{**} \circ u_L \circ F$ uniformly tightens $G$.
\end{proof}

Proposition \ref{Proposition: post-composing continuous maps} is an obvious consequence of the definition of uniform universality.
If $\mathcal{C}$ is the class of all envelopes then it says that post-composing a continuous map with a uniformly universal envelope 
will again yield a uniformly universal envelope.

We immediately obtain effective versions of Propositions 
\ref{Proposition: extending global probe, non-uniform}
and
\ref{Proposition: extending probe, non-uniform}.
We only state an effective version of the latter, as it is less straightforward:

\begin{cor}\label{Corollary: uniform extension probes}
	Let $f \colon X \to Y$ be a function between computable $T_0$ spaces.
	Let $\mathcal{C}$ be a class of approximation spaces.
	Let $F \colon X \to L$ be a uniformly $\mathcal{C}$-universal envelope.
	Let $\widetilde{X}$ and $Z$ be a effective $T_0$ spaces.
	Let $M$ be an effective continuous lattice.
	Let $i \colon Z \to M$ be a continuous map.
	Let $i_* \colon Z^{\widetilde{X}} \to M^{\widetilde{X}}$ be defined by 
	$i_*(\varphi)(\widetilde{x}) = i(\varphi(\widetilde{x}))$.
	Assume that 
	$i_* \colon Z^{\widetilde{X}} \to M^{\widetilde{X}}$ 
	belongs to $\mathcal{C}$.
	Let 
	$\alpha \colon \widetilde{X} \to X$
	and 
	$\beta \colon \widetilde{X} \times Y \to M$
	be continuous functions.
	Let 
	$
		h \colon Y \to M^{\widetilde{X}},
		\;
		h(y)(\widetilde{x}) = i_* \circ \beta(\widetilde{x}, y).
	$
	Then the following are equivalent:
	\begin{enumerate}
		\item 
			Every point 
			$(\widetilde{x}, \alpha(\widetilde{x})) \in \widetilde{X} \times X$ 
			is a point of continuity of the function 
			$\psi \colon \widetilde{X} \times X \to Z, 
			\psi(\widetilde{x}, x) = \beta(\widetilde{x}, f (x))$
		\item 
			The function 
			$E_{L, M^{\widetilde{X}}}(h) \colon L \to M^{\widetilde{X}}$ 
			satisfies 
			\[
				E_{L, M^{\widetilde{X}}}(h)\left(F \circ \alpha(\widetilde{x})\right)(\widetilde{x})
				= 
				\beta(\widetilde{x}, f \circ \alpha (\widetilde{x})).
			\]
	\end{enumerate}
\end{cor}

If the right hand side of a composition is allowed to be the universal envelope of a discontinuous function then the composition may no longer be automatically uniformly universal. 
As mentioned in the introduction, this leads to the study of how functions and their envelopes transport open sets of the co-domain to open sets of the domain.

For an arbitrary function $f \colon X \to Y$, let 
\[ 
	\intr{f} \colon \O(Y) \to \O(X),
	\; 
	\intr{f}(V) = \intr{f^{-1}(V)}.
\]
For a continuous function $F \colon X \to \O^2(Y)$, let 
\[
	\starred{F} \colon \O(Y) \to \O(X),
	\; 
	\starred{F}(V) = 
		\Set{x \in X}{F(x) \ni V}.
\]
If $F$ is an envelope of $f$ then we clearly have $\starred{F} \subseteq \intr{f}$.

\begin{lem}\label{Lemma: uniform universality via star}
	Let $f \colon X \to Y$ be a function between effective $T_0$ spaces.
	Let $F \colon X \to \O^2(Y)$ be an envelope of $f$.
	Then $F$ is uniformly universal if and only if 
	$\starred{F}(V) = \intr{f}(V)$
	for all open sets 
	$V \in \O(Y)$.
\end{lem}
\begin{proof}
	Assume that $F$ is uniformly universal.
	Let $V \in \O(Y)$.
	Then 
	$\starred{F}(V) \subseteq \intr{f^{-1}(V)}$
	since $F$ is an envelope.
	Let $x \in \intr{f^{-1}(V)}$.
	Then $V$ is a robust property of $f(x)$.
	It follows that $F(x) \ni V$.
	Thus, $\starred{F}(V) = \intr{f^{-1}(V)}$.

	Conversely, assume that 
	$\starred{F}(V) = \intr{f}(V)$
	for all $V \in \O(Y)$.
	If $V$ is a robust property of $f(x)$ 
	then $x \in \intr{f^{-1}(V)}$.
	It follows that $V \in F(x)$.
	It follows from Theorem \ref{Theorem: motivation}
	that $F$ is uniformly universal.
\end{proof}

\begin{lem}\label{Lemma: star respects composition}
	Let $X$, $Y$, and $Z$ be effective $T_0$ spaces.
	Let $F \colon X \to \O^2(Y)$
	and $G \colon Y \to \O^2(Z)$
	be continuous maps.
	We have 
	$\starred{(G \bullet F)} = \starred{F} \circ \starred{G}$.
\end{lem}
\begin{proof}
	One easily verifies that 
	$\starred{F} = F^* \circ \nu_{\O(Y)}$.
	Using this, we obtain
	\[
		\starred{(G \bullet F)}
		=
		(\nu_{\O(Z)}^* \circ G^{**} \circ F)^* \circ \nu_{\O(Z)}
		=
		F^* \circ G^{***} \circ \nu_{\O(Z)}^{**} \circ \nu_{\O(Z)}.
	\]
	Now, by naturality of the unit $\nu_{\O(Z)}$ we have  
	$\nu_{\O(Z)}^{**} \circ \nu_{\O(Z)} = \nu_{\O^3(Z)} \circ \nu_{\O(Z)}$,
	so that 
	\[
		\starred{(G \bullet F)}
		=
		F^* \circ G^{***} \circ \nu_{\O^3(Z)} \circ \nu_{\O(Z)}
		=
		F^* \circ \nu_{\O(Y)} \circ G^* \circ \nu_{\O(Z)}
		= 
		\starred{F} \circ \starred{G}.
	\]
	The second-to-last equality is again using the naturality of $\nu$.
\end{proof}

We conclude:

\begin{thm}\label{Theorem: characterisation of universality for general envelopes}
	Let $f \colon X \to Y$ and $g \colon Y \to Z$ be functions between computable $T_0$ spaces.
	Let $F \colon X \to \O^2(Y)$ and $G \colon X \to \O^2(Z)$ be the principal $\O^2$-envelopes of $f$ and $g$ respectively.
	Assume that $F$ and $G$ are uniformly universal.
	Then $G \bullet F$ is a uniformly universal envelope of $g \circ f$ if and only if 
	$\intr{(g \circ f)} = \intr{f} \circ \intr{g}$.
\end{thm}

Theorem \ref{Theorem: characterisation of universality for general envelopes} leads to the question 
for which pairs of functions $f \colon X \to Y$ and $g \colon Y \to Z$ we have the equality
$\intr{(g \circ f)} = \intr{f} \circ \intr{g}$.
We obtain a result that is sharp in a certain sense:

\begin{thm}\label{Theorem: interior inverse and openness}
	\hfill 
	\begin{enumerate}
		\item 
		Let $f \colon X \to Y$ and $g \colon Y \to Z$ be functions between effective $T_0$ spaces.
		Then we have $\intr{f} \circ \intr{g} \subseteq \intr{(g \circ f)}$.
		\item 
		Let $f \colon X \to Y$ be a function between effective $T_0$ spaces.
		Let $Z$ be an effective $T_0$ space with at least two points.
		We have 
		$\intr{f} \circ \intr{g} = \intr{(g \circ f)}$
		for all 
		$g \colon Y \to Z$
		if and only if 
		$f$ 
		sends open sets to open sets.
		\item 
		Let $g \colon Y \to Z$ be a function between effective $T_0$ spaces.
		Let $X$ be an effective $T_0$ space.
		We have 
		$\intr{f} \circ \intr{g} = \intr{(g \circ f)}$
		for all 
		$f \colon X \to Y$
		if and only if 
		$g$ 
		is continuous.
	\end{enumerate}
\end{thm}
\begin{proof}
	\hfill
	\begin{enumerate}
		\item Obvious.
		\item 
		Assume that $f$ sends open sets to open sets.
		Let $V \in \O(Z)$.
		Since we always have 
		$\intr{f} \circ \intr{g}(V) \subseteq \intr{(g \circ f)}(V)$
		it remains to prove the converse inclusion.
		Let $x \in \intr{(g \circ f)}(V)$.
		Since $f$ maps open sets to open sets,
		the set 
		$f(\intr{(g \circ f)}(V))$
		is contained in 
		$\intr{g}(V)$.
		By applying $f^{-1}$ to both sides of this containment we obtain 
		\[
			f^{-1}(\intr{g}(V)) \supseteq \intr{\intr{(g \circ f)}(V)} \ni x.
		\] 
		Thus, $x$ is contained in an open set which is contained in 
		$f^{-1}(\intr{g}(V))$.
		It follows that $x \in \intr{f^{-1}(\intr{g}(V))} = \intr{f} \circ \intr{g}(V)$.

		This establishes the ``if'' part of the claim.
		Let us now establish the ``only if'' part.
		Fix $t, b \in Z$ with $t \neq b$ and $b \not \geq t$ in the specialisation order of $Z$.
		Let $U \in \O(X)$.
		Our aim is to show that $f(U)$ is an open subset of $Y$.
		Consider the function 
		\[
			g \colon Y \to Z,
			\; 
			g(y) = 
			\begin{cases}
				t &\text{if }y \in f(U) \\ 
				b &\text{if }y \notin f(U).
			\end{cases}
		\]
		By assumption on $t$ and $b$ there exists an open set $V \in \O(Z)$ such that 
		$t \in V$ and $b \notin V$.
		We have by assumption on $f$:
		\[
			U \subseteq 
			\intr{f^{-1}(f(U))}
			= \intr{(g \circ f)}(V)
			= \intr{f} \circ \intr{g}(V)
			= \intr{f^{-1}(\intr{f(U)})}
			\subseteq f^{-1}(\intr{f(U)}).
		\]
		Applying $f$ to both sides of the containment we find that 
		$f(U) \subseteq \intr{f(U)}$,
		hence 
		$f(U) = \intr{f(U)}$.
		\item 
			If $g$ is continuous, then $\intr{g} = g^{-1}$, so that 
			$\intr{(g \circ f)} = \intr{f} \circ \intr{g}$ follows immediately.

			Conversely, assume that 
			$g$ 
			satisfies 
			$\intr{(g \circ f)} = \intr{f} \circ \intr{g}$
			for all functions  $f \colon X \to Y$.
			Let $V \in \O(Z)$.
			We claim that $g^{-1}(V)$ is open.
			If $g^{-1}(V)$ is empty then the claim is trivial.
			Thus, assume that $g^{-1}(V)$ is non-empty.
			Let $y \in g^{-1}(V)$.
			Let $f \colon X \to Y$ be the constant map with value $y$.
			Then, by assumption,
			\[
				X
				=
				\intr{f^{-1}(g^{-1}(V))}
				=
				\intr{f^{-1}(\intr{g^{-1}(V)})}
				\subseteq 
				f^{-1}(\intr{g^{-1}(V)})
			\]
			It follows that 
			$f^{-1}(\intr{g^{-1}(V)}) = X$,
			so that 
			$y \in \intr{g^{-1}(V)}$.
			Thus, $g^{-1}(V) = \intr{g^{-1}(V)}$,
			so that $g^{-1}(V)$ is open.\qedhere
	\end{enumerate}
\end{proof}

As an obvious corollary to Theorems \ref{Theorem: characterisation of universality for general envelopes} and \ref{Theorem: interior inverse and openness}
we obtain that the composition of two uniformly universal envelopes $G$ of $g$ and $F$ of $f$ is uniformly universal if the function $f$ sends open sets to open sets.
Conversely, observing that the functions $g$ used in the proof of Theorem \ref{Theorem: interior inverse and openness}.2 are uniformly envelopable, we obtain that if 
$G \bullet F$ 
is uniformly universal for all uniformly universal envelopes 
$G \colon Y \to \O^2(Z)$,
where $Z$ is a fixed space containing at least two points,
then $f$ sends open sets to open sets.

In this sense our result is sharp:
openness of $f$ is sufficient and necessary in order for \emph{all} compositions 
$G \bullet F$ 
to be uniformly envelopable again.

Of course, the function $f$ need not be open in order for the equation 
$\intr{(g \circ f)} = \intr{f} \circ \intr{g}$ 
to hold for \emph{some} $g$.
A stronger result would be obtained if one could characterise for a given function 
$f \colon X \to Y$ 
and a given space 
$Z$
precisely those functions 
$g \colon Y \to Z$ 
that satisfy 
$\intr{(g \circ f)} = \intr{f} \circ \intr{g}$.

A result of this generality seems to be completely out of reach, however.
Already for the -- continuous -- diagonal embedding 
$f \colon \C \to \C \times \C$
we can find a single function 
$g \colon \C \times \C \to \{0,1\}$
such that the question whether 
$\intr{(g \circ f)} = \intr{f} \circ \intr{g}$
is equivalent to a long-standing open problem in complex dynamics.

The following argument will make heavy use of definitions and results from complex dynamics that due to space restrictions we can only briefly summarize. 
The results we use are well known, but not easy to prove.
See \text{e.g.} \cite{CarlesonGamelin, MilnorDynamics} for good introductions to complex dynamics and \cite{BrannerMandelbrot} for a good introduction to the Mandelbrot set.
See also \cite{ComputabilityJulia, HertlingMandelbrot} for results on the computability of Julia sets and the Mandelbrot set.

Let $X$ be a set.
The orbit of a point $x \in X$ under a function $f \colon X \to X$
is the set $\Set{f^{(n)}(x)}{n \in \N}$.

The Mandelbrot set 
$M \subseteq \C$ 
is the set of points 
$c \in \C$ 
such that the orbit of 
$0 \in \C$ 
under the polynomial 
$z \mapsto z^2 + c$
is bounded.

The filled-in Julia set $K_c$ of a point $c \in \C$ is the set of all points $z \in \C$
such that the orbit of $z$ under the polynomial $z^2 + c$ is bounded.
The Julia set $J_c$ of a point $c \in \C$ is the boundary $\partial K_c$ of the filled-in Julia set.

From these definitions we immediately obtain that $c \in M$ if and only if $c \in K_c$.
Letting $\chi_M \colon M \to \{0,1\}$ denote the characteristic function of the Mandelbrot set,
we obtain the commutative diagram:

\begin{center}
    \begin{tikzcd}
        \C \times \C \arrow{r}{\Phi} & \{0,1\} \\ 
        \C \arrow{u}{x \mapsto (x,x)}  \arrow[swap]{ur}{\chi_M}& 
    \end{tikzcd}
\end{center}

where 
$u \colon \C \to \C \times \C$
is the diagonal embedding and 
$\Phi(z,c) = 1$ 
if and only if 
$z \in K_c$.
It is easy to see that all functions with finite co-domain are uniformly envelopable.
See also Theorem \ref{Theorem: Noetherian} below for a sharper result.
It follows that $\Phi$, $\chi_M$, and $\Phi \circ u$ are uniformly envelopable.
Let $\widetilde{\chi_M}$ denote the principal $\O^2(\{0,1\}))$-envelope of $\chi_M$.
Let 
$\widetilde{\Phi} \colon \C \times \C \to \O^2(\{0,1\})$
denote the principal $\O^2(\{0,1\}))$-envelope of $\Phi$.
The function $u$ is continuous and hence uniformly envelopable with 
uniformly universal envelope 
$\nu_{\C \times \C} \circ u \colon \C \to \O^2(\C \times \C)$.
Clearly, 
$\widetilde{\Phi} \bullet (\nu_{\C \times \C} \circ u) = \widetilde{\Phi} \circ u$.
We will show that 
$\widetilde{\Phi} \circ u$ is the principal 
$\O^2(\{0,1\})$-envelope of $\Phi \circ u$ if and only if the so-called hyperbolicity conjecture holds true.
A parameter $c \in \C$ is called \emph{hyperbolic} if the polynomial $p_c(z) = z^2 + c$ has an attracting 
cycle, see \textit{e.g.} \cite[pp. 127--128]{CarlesonGamelin}, \cite[p. 82 ff.]{BrannerMandelbrot}.
The set of hyperbolic parameters is an open subset of $M$.
The hyperbolicity conjecture \cite[p. 83]{BrannerMandelbrot} states that the set of hyperbolic parameters is equal to the interior of $M$.
It is considered to be one of the major unsolved open problems in complex dynamics.

We clearly have:
\[
    \widetilde{\chi_M}(c)
    =
    \begin{cases}
        \nu_{\{0,1\}}(1) &\text{if }c \in \intr{M}, \\
        \nu_{\{0,1\}}(0) &\text{if }c \in M^C, \\ 
        \emptyset        &\text{if }c \in \partial M.
    \end{cases}
\]

By \cite{DouadyContinuous} the sets 
$K_c$
and  
$J_c$
both depend continuously on $c$ in the Hausdorff metric when $c$ ranges over the set 
$\C \setminus \partial M$.
It follows that given $(z,c) \in \C^2$ we can test (relative to some oracle) if $c \in \C \setminus \partial M$
and then semi-decide relative to some oracle if $c \in K_c \setminus J_c$.
Again by \cite{DouadyContinuous} the predicate $z \notin K_c$ is open for all $(z,c) \in \C$.
There are two cases not covered by these (relative) semi-decision procedures:
the case where $z \in \partial K_c$ and the case where $c \in \partial M$.
But in both cases the value of $\Phi(z,c)$ changes under arbitrarily small perturbations.

In summary, we have:
\[
    \widetilde{\Phi}(z,c)
    =
    \begin{cases}
        \nu_{\{0,1\}}(0) &\text{if } z \notin K_c, \\
        \nu_{\{0,1\}}(1) &\text{if } c \in \intr{M} \text{ and } z \in \intr{K_c},\\
        \emptyset        &\text{otherwise.}
    \end{cases}
\]

Now, 
\[
    \widetilde{\Phi}(c,c) 
    =
    \begin{cases}
        0 &\text{if } c \notin M, \\
        1 &\text{if } c \in \intr{M} \text{ and } z \in \intr{K_c},\\
        \bot &\text{otherwise.}
    \end{cases}
\] 

It is shown in \cite[Proposition 5]{Orsay} that if $c$ belongs to a hyperbolic component of the interior of $M$, then $c$ belongs to the interior of $K_c$.
Conversely, assume that $c$ belongs to a non-hyperbolic component of the interior of $M$.
Then it follows from the Fatou-Sullivan classification 
\cite[Theorem 1.3, Theorem 2.1]{CarlesonGamelin}
that $K_c$ has empty interior, so that $c \in \partial K_c$.

It follows that $\widetilde{\Phi}(c,c)$ is equal to $\widetilde{\chi_M}(c)$ if and only if the interior of $M$ is the union of the hyperbolic components.

\section{A connection with Noetherian spaces}

Since uniform universality is a very desirable property it seems natural to ask if there exist simple criteria on a function 
$f \colon X \to Y$ 
that a-priori guarantee uniform universality. 
Perhaps the simplest conceivable criteria are those that put constraints on the domain or co-domain and let $f$ be arbitrary otherwise.

Recall that a topological space is Noetherian if all descending sequences of closed sets stabilise after finitely many steps.
Noetherian spaces admit the following, straightforwardly verified, equivalent characterisation: 
a space $Y$ is Noetherian if and only if all open subsets of $Y$ are compact, cf.~\cite{Goubault-Larrecq, Noetherian}.

\begin{thm}\label{Theorem: Noetherian}
    The following are equivalent for an effective $T_0$ space $Y$:
    \begin{enumerate}
        \item For all effective $T_0$ spaces $X$, all functions $f \colon X \to Y$ are uniformly envelopable.
        \item All functions $f \colon 2^{\N} \to Y$ are uniformly envelopable.
        \item $Y$ is Noetherian.
    \end{enumerate}
\end{thm}
\begin{proof}
    Let us first show the implication $(3) \Rightarrow (1)$.
    Assume that $Y$ is Noetherian. 
    Then every open subset of $Y$ is compact.
    Let $f \colon X \to Y$ be any function.
    Let $x \in X$.
    Let $U \in \O(Y)$ be a robust property of $f(x)$.
    Then the set 
    $\ucl{U} = \Set{V \in \O(Y)}{V \supseteq U}$
    is open.
    Let $W \in \O(X)$ be such that $x \in W \subseteq f^{-1}(U)$.
    Consider the $\O^2(Y)$-envelope 
    \[
        G(z) 
        =
        \begin{cases}
            \ucl{U}   &\text{if }z \in W, \\ 
            \emptyset &\text{otherwise.} 
        \end{cases}
    \]
    This is an envelope of $f$ by construction.
    It satisfies $U \in G(x)$.
    It follows that the principal $\O^2(Y)$-envelope $F$ satisfies $U \in F(x)$.
    Since $U$ was an arbitrary robust property it follows from Theorem \ref{Theorem: motivation} that $F$ is uniformly universal.

    The implication $(1) \Rightarrow (2)$ is trivial.
    It thus remains to show the implication $(2) \Rightarrow (3)$.
    Assume that all functions $f \colon 2^{\N} \to Y$ are uniformly envelopable.
    Let $U \in \O(Y)$ be an open subset of $Y$.
    We will show that $U$ is compact.
    We may assume that $U$ contains at least two points, for otherwise the claim is trivial.
    Let $0 \in 2^{\N}$ denote the constant zero sequence.
    It suffices to show that there exists a function 
    $f \colon 2^{\N} \to Y$ 
    such that the set of robust properties of 
    $f(0)$ 
    is 
    $\Set{V \in \O(Y)}{V \supseteq U}$.
    Indeed, since the principal $\O^2(Y)$-envelope $F$ of $f$ is assumed to be uniformly universal
    we must have that $F(0)$ is the filter of robust properties of $f(0)$.
    It then follows that $\Set{V \in \O(Y)}{V \supseteq U}$ is open, \textit{i.e.}, $U$ is compact.

    To prove the claim, fix $y_0 \in U$. 
    Note that $U \setminus \{y_0\}$ is non-empty by assumption on $U$.
    Observe that there exist continuum-many disjoint sequences in $2^{\N}$ that converge to $0 \in 2^{\N}$. 
    Since $U \setminus \{y_0\}$ contains at most continuum-many points, we can find a set of distinct points 
    $p_{y,n} \in 2^{\N}$
    where $y$ ranges over $U \setminus \{y_0\}$ 
    and $n$ ranges over $\N$,
    such that for all $y \setminus \{y_0\}$ we have that $p_{y,n} \to 0$ as $n \to \infty$.
    Thus, we obtain a well-defined partial function $f$ on $2^{\N}$ by letting 
    $f(p_{y,n}) = y$.
    Let $f$ map all remaining points of $2^{\N}$ to $y_0$.

    Then $f^{-1}(U) = 2^{\N}$, so that $U$ is a robust property of $f(0)$.
    Now, let $V \in \O(Y)$ be an open set
    not containing $U$
    with $y_0 \in V$.
    Then there exists $y \in U$ such that $y \notin V$.
    It follows that 
    $f^{-1}(V) \subseteq 2^{\N} \setminus \Set{p_{y,n}}{n \in \N}$ 
    which implies that $f^{-1}(V)$ cannot be a neighbourhood of $0$.
    It follows that $V$ is not a robust property of $f(0)$.
    This finishes the proof.
\end{proof}

While being Noetherian is a rather strong restriction, there are some interesting non-trivial examples of Noetherian effective $T_0$ spaces.
Notably the spectra of Noetherian commutative rings with the Zariski topology, provided that their topology is countably based,
and well-quasi orders on $\N$ with the Alexandroff topology \cite{Noetherian,Goubault-Larrecq}.

Let us now turn to the analogous question for $X$.
We first require a technical lemma:

\begin{lem}\label{Lemma: spaces w/o infinite compact sets}
    Let $X$ be an effective $T_0$ space that does not contain an infinite compact set.
    Then $X$ is countable, countably based, and for all $x \in X$ the set $\ucl{x}$ is open and finite.
\end{lem}
\begin{proof}
    Let $x \in X$.
    Let $(x_n)_n$ be a sequence that converges to $x$.
    Then the set $\Set{x_n}{n \in \N} \cup \{x\}$ is compact, and hence finite.
    It follows that $x_n \in \ucl{x}$ for all sufficiently large $n$.
    Since the topology on $X$ is sequential, the set $\ucl{x}$ is open.
    Since $\ucl{x}$ is compact, it is finite.

    It remains to show that $X$ is countable.
    Then $X$ is countably based, for the sets $\ucl{x}$, where $x \in X$, form a basis of the topology.
    Since $X$ is an effective $T_0$ space, it admits a countable pseudobase $(B_n)_n$ \cite[Theorem 13]{SchroederAdmissibility}.
    Let $y \in \ucl{x}$. Then, by definition (\cite[Section 3.1]{SchroederAdmissibility}), there exists $j \in \N$ with 
    $y \in B_j \subseteq \ucl{x}$.
    Since $\ucl{x}$ is finite, there exist finitely many numbers $j_1,\dots,j_s \in \N$ such that 
    $\ucl{x} = B_{j_1}\cup\dots\cup B_{j_s}$.
    Fix any map that sends $x \in X$ to $\langle j_1,\dots,j_s\rangle \in \N^*$
    with $\ucl{x} = B_{j_1}\cup\dots\cup B_{j_s}$.
    Since $X$ is $T_0$, this map is injective. 
    It follows that $X$ is countable.
    Thus, everything is shown.
\end{proof}

\begin{thm}\label{Theorem: spaces X such that all f with domain X are uniformly envelopable}
    Let $X$ be an effective $T_0$ space.
    Then all functions $f \colon X \to Y$ are uniformly envelopable if and only if $X$ contains no infinite compact sets.
\end{thm}
\begin{proof}
    Assume that $X$ contains an infinite compact set.
    Then $X$ contains an infinite sequence $(x_n)_n$ of distinct points that 
    converges to a limit $x_{\infty}$ with $x_{\infty} \neq x_n$ for all $n$.
    Let $f \colon X \to \R$ be defined as follows:
     $f(x_n) = 1/n$ 
    and $f(x) = 1$ 
    for all $x \in X \setminus \Set{x_n}{n \in \N}$.
    Let $F \colon X \to \Kb(\R)$ be the principal $\Kb(\R)$-envelope of $f$.
    Then $F(x_{\infty}) \ni 0 = \lim_{n \to \infty} f(x_n)$.
    But $f^{-1}((0,+\infty)) = X$, so that the interval $(0,+\infty)$ is a robust property of $f(x_{\infty})$
    wich is not witnessed by $F$.
    It follows that $F$ is not uniformly universal.

    Assume that $X$ contains no infinite compact subsets.
    Let $f \colon X \to Y$ be any function. 
    By Lemma \ref{Lemma: spaces w/o infinite compact sets}
    the space $X$ contains at most countably many points
    and
    for all $x \in X$ the set $\ucl{x}$ is a finite open set.
    It follows in particular that the set $f(\ucl{x})$ is a finite and hence compact subset of $Y$.

    There hence exist continuous functions 
    $
        \omega \colon \N \to \O(X)
    $
    and
    $
        \varphi \colon \N \to \K(Y),
    $
    such that $\omega$ maps $\N$ onto the set of all sets of the form $\ucl{x}$ for $x \in X$,
    and $\varphi(n) = f(\omega(n))$.

    Let 
    $
        F(x) = \Set{V \in \O(Y)}{V \supseteq f(\ucl{x})}.
    $
    Then $F$ is continuous, for it is computable relative to the oracles $\omega$ and $\varphi$ above.
    Indeed, consider the following algorithm:
    given 
    $x \in X$ 
    and 
    $V \in \O(Y)$
    search for a number $n \in \N$ such that 
    $\omega(n) \supseteq \ucl{x}$
    and 
    $\varphi(n) \subseteq V$.
    Halt if and only if such a number exists.
    On the one hand, if this algorithm halts then there exists $n \in \N$ such that 
    $
        f(\ucl{x}) \subseteq f(\omega(n)) = \varphi(n) \subseteq V.
    $
    On the other hand, if $f(\ucl{x}) \subseteq V$ then the algorithm will halt, 
    for there exists $n \in \N$ with $\omega(n) = \ucl{x}$.

    Now, $F(x) \subseteq \nu_Y \circ f(x)$ by definition.
    The function $F$ is hence an envelope of $f$.
    If $V \in \O(Y)$ is a robust property of $f(x)$ then clearly $V \supseteq f(\ucl{x})$.
    It follows that $F$ is uniformly universal.
\end{proof}

\section{Regular envelopes}

\begin{defi}\label{Definition: separated/regular}
    An approximation space $\xi_L \colon Y \to L$ is called \emph{separated}
    if the map 
    \[
        \xi_L^{-1}(\ucl{\cdot}) \colon L \to \A(Y),
        \;
        \xi_L^{-1}(\ucl{x}) = \Set{y \in Y}{\xi_L(y) \geq x}
    \]
    is well-defined and continuous.
    It is called \emph{computably separated} if $\xi_L^{-1}(\ucl{\cdot})$ is computable.

    A separated approximation space 
    $\xi_L \colon Y \to L$ 
    is called \emph{regular}
    if for all 
	$x \in L$ and all $V \in \O(L)$ with $x \in V$ 
    there exists $U \in \O(L)$ with $x \in U$ and
	$\clos{\xi_L^{-1}(U)} \subseteq \xi_L^{-1}(V)$.
\end{defi}

Note that the definition of separated approximation space given here is strictly more general than the one 
given in \cite{NeumannPhD}.
The name ``separated'' is motivated by the following observation:

\begin{prop}
	Let $X$ be an effective $T_0$ space.
	Then $X$ is (computably) Hausdorff if and only if the approximation space 
	$\nu_X \colon X \to \O^2(X)$
	is (computably) separated.
\end{prop}

The name ``regular'' is motivated by the following observation:

\begin{prop}\label{Proposition: O^2(X) regular -> X regular}
	Let $X$ be a an effective $T_0$ space.
	If the approximation space 
	$\nu_X \colon X \to \O^2(X)$
	is regular, then $X$ is a regular topological space.
\end{prop}

For countably based spaces the converse of Proposition \ref{Proposition: O^2(X) regular -> X regular} holds true as well:

\begin{prop}\label{Proposition: Kb(Y) regular}
	Let $Y$ be a complete effective metric space. 
	Then the approximation spaces 
	$(\kappa_\bot)_Y \colon Y \to \Kb(Y)$ 
	and 
	$\nu_Y \colon Y \to \O^2(Y)$ 
	are regular.
\end{prop}
\begin{proof}
	The space $\Kb(Y)$ is an approximation space since $\Kb(Y)$ is injective by \cite[Proposition 3.24]{NeumannPhD}.
	It is easy to see that $\Kb(Y)$ is separated.
	Let us now verify that the approximation space is regular.
	The element $\bot$ is contained in a single open set, namely $\Kb(Y)$.
	We have 
	\[ 
		(\kappa_\bot)_Y^{-1}(\ucl{\bot}) = Y = \clos{Y} = (\kappa_\bot)_Y^{-1}(\Kb(Y)). 
	\]
	Now, let $K \in \K(Y)$ and let $V \in \O(\Kb(Y))$ with $K \in V$.
	The sets of the form 
	\[
		[U] 
		= 
		\Set{K \in \K(Y)}
			{K \subseteq U},
	\]
	where $U \in \O(Y)$,
	form a basis for the topology of $\K(Y)$ \cite[Proposition 2.40]{NeumannPhD}.
	There hence exists $U \in \O(Y)$ with $K \in [U] \subseteq V$.
	We have $(\kappa_\bot)_Y^{-1}(\ucl{K}) = K$
	and $(\kappa_\bot)_Y^{-1}([U]) = U$.
	Since $K$ is compact and $Y$ is regular Hausdorff there exists a set 
	$W \in \O(Y)$ such that 
	\[ 
		K \subseteq W \subseteq \clos{W} \subseteq U \subseteq (\kappa_\bot)_Y^{-1}(V).
	\] 
	We have $W = (\kappa_\bot)_Y^{-1}([W])$ with $K \in [W]$ and
	\[
		(\kappa_\bot)_Y^{-1}(\ucl{K}) \subseteq (\kappa_\bot)_Y^{-1}([W]) \subseteq \clos{(\kappa_\bot)_Y^{-1}([W])} \subseteq (\kappa_\bot)_Y^{-1}(V).
	\] 
	This proves the claim.

	Now consider the approximation space $\nu_Y \colon Y \to \O^2(Y)$.
	By \cite[Theorem 6.11]{DeBrechtKawai} a basis for the open sets of $\O^2(Y)$ is given by sets the of the form 
	\[
		[U_1,\dots,U_n]
		=
		\Set{\mathcal{U} \in \O^2(Y)}
			{U_1 \in \mathcal{U} \land \dots \land U_n \in \mathcal{U}}.
	\]
	with $U_1,\dots,U_n \in \O(Y)$.
	Thus, 
	let 
	$\mathcal{U} \in \O^2(Y)$ 
	and assume that 
	$\mathcal{U} \in [U_1,\dots,U_n]$.
	Since Polish spaces are consonant (see, \textit{e.g.}, \cite[Section 6]{DeBrechtKawai} or  \cite[Section 2.6]{NeumannPhD}) there exists a family of compact sets 
	$(K_i)_{i \in I}$
	such that 
	$U \in \mathcal{U}$
	if and only if 
	$U \supseteq K_i$
	for some $i \in I$.
	It follows that there exist $i_1,\dots, i_n$ such that 
	$K_{i_j} \subseteq U_j$
	for $j = 1,\dots,n$.
	Again using that $Y$ is a regular topological space, we can find for each $j = 1,\dots,n$ an open set $V_j \in \O(Y)$
	with 
	$K_{i_j} \subseteq V_j \subseteq \clos{V_j} \subseteq U_j$.
	Then 
	$\mathcal{U} \in [V_1,\dots,V_n]$
	and 
	\begin{align*}
		\nu_Y(\ucl{\mathcal{U}}) 
		&\subseteq \nu_{Y}([V_1,\dots,V_n]) 
		= V_1 \cap \dots \cap V_n \\
		&\subseteq 
		\clos{V_1 \cap \dots \cap V_n}
		\subseteq 
		U_1 \cap \dots \cap U_n
		=
		\nu_Y^{-1}([U_1 \cap \dots \cap U_n]).
    \end{align*}
	This proves the claim.
\end{proof}

A straightforward compactness argument shows that the interpolation property of regular approximation spaces 
extends to all compact sets:

\begin{lem}\label{Lemma: regular interpolation property compacts}
	A separated approximation space $\xi_L \colon Y \to L$ is regular 
	if and only if 
	for all 
	$K \in \K(L)$ and all $V \in \O(L)$ with $K \subseteq V$ 
    there exists $U \in \O(L)$ with $K \subseteq U$ and
	$\clos{\xi_L^{-1}(U)} \subseteq \xi_L^{-1}(V)$.
\end{lem}

\begin{prop}\label{Proposition: regular exponential}
		Let 
		$\xi_L \colon Y \to L$ 
		be a regular approximation space.
		Let $Z$ be an effective $T_0$ space.
		Assume that $L^Z$ carries the compact-open topology.
		Then 
		\[ 
			(\xi_L)_* \colon Y^Z \to L^Z,
			\; 
			(\xi_L)_*(h)(z) = \xi_L(h(z))
		\]
		is again a regular approximation space.
\end{prop}
\begin{proof}
		Let us first show that 
		$(\xi_L)_* \colon Y^Z \to L^Z$
		is separated.
		For all 
		$\varphi \colon Z \to L$
		we have 
		\[
			(\xi_L)_*^{-1}(\ucl{\varphi})
			= 
			\Set{h \colon Z \to Y}
				{\forall z \in Z. h(z) \in \xi_L^{-1}(\ucl{\varphi(z)})}.
		\]
		By assumption, the map 
		$
			\xi_L^{-1}(\ucl{\cdot}) \colon L \to \A(Y)
		$
		is well-defined and continuous.
		It follows that the predicate 
		$\exists z \in Z. h(z) \notin \xi_L^{-1}(\ucl{\varphi(z)})$
		is relatively semi-decidable in $\varphi$ and $h$.
		This proves the claim. 

		Now let us show that the approximation space is regular.
		Let $\varphi \colon Z \to L$.
		Let $\varphi \in V \in \O(L^Z)$.
		By assumption $L^Z$ carries the compact-open topology,
		so that we may assume without loss of generality that 
		$V = \Set{\psi \colon Z \to L}{\psi(K) \subseteq W}$
		for some $K \in \K(Z)$ and $W \in \O(L)$.
		Thus, 
		$\varphi(K) \subseteq W$.
		By Lemma \ref{Lemma: regular interpolation property compacts} 
		there exists a set $U \in \O(L)$ with 
		$\varphi(K) \subseteq U$
		and 
		$\clos{\xi_L^{-1}(U)} \subseteq \xi_L^{-1}(W)$.
		We thus have 
		$\varphi \in \Set{\psi \colon Z \to L}{\psi(K) \subseteq U}$
		and
		\begin{align*}
			(\xi_L)_*^{-1}\left(\Set{\psi \colon Z \to L}{\psi(K) \subseteq U}\right)
			&= \Set{h \colon Z \to Y}{h(K) \subseteq \xi_L^{-1}(U)}\\
			&\subseteq \clos{\Set{h \colon Z \to Y}{h(K) \subseteq \xi_L^{-1}(U)}}\\
			&\subseteq \Set{h \colon Z \to Y}{h(K) \subseteq \xi_L^{-1}(W)}\\
			&= (\xi_L)_*^{-1}(V).
		\end{align*}
		This proves the claim.
\end{proof}

\begin{prop}\label{Proposition: examples where exponential has compact-open topology}
	Let $L$ and $Z$ be countably based effective $T_0$ spaces.
	If $Z$ is locally compact 
	or 
	$L$ is $\omega$-continuous and $Z$ is Polish,
	then $L^Z$ carries the compact-open topology.
\end{prop}
\begin{proof}
	It follows from Theorem 7 and Section 4.4 in \cite{SchroederAdmissibility} that the space $L^Z$ carries the sequentialisation of the compact-open topology.
	If $Z$ is locally compact and $L$ is countably based, 
	then the compact-open topology on 
	$L^Z$ is countably based.
	Thus, the compact-open topology agrees with its sequentialisation.
	This establishes the first part of the claim.

	To prove the second part, observe that the topology on $L^Z$ is finer than the compact-open topology and coarser than the Scott topology
	(cf.~\cite[Proposition 3.9]{NeumannPhD}).
	It hence suffices to show that the Scott topology on $L^Z$ coincides with the compact-open topology.
	This can be proved by a rather straightforward adaptation of the proof of Theorem 4.1 in \cite{Consonant}, which establishes the result in the special case 
	where $L = \Sigma$.
	For the sake of completeness, the proof is included in Appendix \ref{Appendix: Proof of compact-open proposition}.
\end{proof}

Propositions \ref{Proposition: regular exponential} and \ref{Proposition: examples where exponential has compact-open topology} show that the definition of ``separated approximation space'' chosen here 
is much more general than the one given in \cite[Definition 3.11, Definition 4.1]{NeumannPhD},
since the ``separated'' approximation spaces in \cite{NeumannPhD} 
are only closed under taking exponentials of the form $L^X$ with $X$ compact Hausdorff, see \cite[Proposition 3.12]{NeumannPhD}.

\begin{prop}\label{Proposition: regular approximation spaces closed under pullbacks}
	Let $Y$ and $Z$ be effective $T_0$ spaces.
	Let $g \colon Y \to Z$ be a continuous map.
	Let $\xi_M \colon Z \to M$ be a regular approximation space of $Z$.
	Then 
	$\xi_M \circ g \colon Y \to M$ is a regular approximation space of $Y$.
\end{prop}
\begin{proof}
	Since $g$ is continuous, 
	the map $g^{-1} \colon \A(Z) \to \A(Y)$
	is well-defined and continuous.
	For all $x \in M$ we have 
	$(\xi_M \circ g)^{-1}(\ucl{x}) = g^{-1}(\xi_M^{-1}(\ucl{x}))$.
	Separatedness follows.

	Let $x \in M$ and $x \in V \in \O(M)$.
	Then by regularity of $M$ there exists $U \in \O(M)$ such that $x \in U \subseteq V$ and 
	\[
		\xi_M^{-1}(\ucl{x}) \subseteq \xi_M^{-1}(U) \subseteq \clos{\xi_M^{-1}(U)} \subseteq \xi_M^{-1}(V).
	\]
	Applying $g^{-1}$ to this we obtain:
	\[
		g^{-1}(\xi_M^{-1}(\ucl{x})) 
		\subseteq g^{-1}(\xi_M^{-1}(U)) 
		\subseteq \clos{g^{-1}(\xi_M^{-1}(U)) }
		\subseteq g^{-1}(\clos{\xi_M^{-1}(U)}) 
		\subseteq g^{-1}(\xi_M^{-1}(V)).
	\]
	This proves the claim.
\end{proof}

Let $\mathcal{R}$ denote the class of all regular approximation spaces.
Accordingly an envelope is called an $\mathcal{R}$-envelope if it takes values in a regular approximation space.
It is called (uniformly) $\mathcal{R}$-universal if it (uniformly) tightens all $\mathcal{R}$-envelopes.

$\mathcal{R}$-universal envelopes witness all robust properties ``up to a small perturbation''
in the following sense:

\begin{prop}
	Let $f \colon X \to Y$ be a function between computable $T_0$ spaces.
	Assume that $Y$ is a normal topological space.
	Let $F \colon X \to L$ be an $\mathcal{R}$-universal envelope of $f$.
	Let $U$ be a robust property of $f(x)$.
	Then $F$ witnesses all robust properties $V \supseteq \clos{U}$.
\end{prop}
\begin{proof}
	Let $W$ be an open neighbourhood of $x$ with $W \subseteq f^{-1}(U)$.
	Let $V \in \O(Y)$ be an open set with $V \supseteq \clos{U}$.
	Our aim is to prove that there exists $O \in \O(L)$ with $F(x) \in O$ and $\xi_L^{-1}(O) \subseteq V$.

	By Urysohn's lemma there exists a continuous function $g \colon Y \to \R$ with 
	$g(y) = 0$ for all $y \in \clos{U}$ and $g(y) = 1$ for all $y \in Y \setminus V$.
	Let $G \colon X \to \Kb(\R)$ be the principal $\Kb(\R)$-envelope of $g \circ f$.
	By Proposition \ref{Proposition: regular approximation spaces closed under pullbacks} 
	$G$ can be viewed as a regular envelope $g^*G$ of $f$
	with values in the approximation space $\kappa_{\bot} \circ g \colon Y \to \Kb(\R)$.
	It follows that $F$ tightens $g^*G$ via a map $\Phi \colon L \to \Kb(\R)$.

	On $W$ the function $g \circ f$ is constant with value $0$.
	It follows that $G$ agrees with $\kappa_\bot \circ g \circ f$ on $W$.
	In particular we have 
	$G(x) \subseteq \left(-\infty, \tfrac{1}{2}\right)$.
	Since $F$ tightens $G$ via $\Phi$ it follows that 
	\[
		F(x) \in \Phi^{-1}\left( \Set{K \in \Kb(\R)}{K \subseteq \left(-\infty, \tfrac{1}{2}\right)} \right).
	\]
	We have $\xi_L^{-1} \circ \Phi^{-1} \leq g^{-1} \circ \kappa_\bot^{-1}$ by definition of the tightening relation.
	Hence:
	\begin{align*}
		\xi_L^{-1} \circ \Phi^{-1}\left( \Set{K \in \Kb(\R)}{K \subseteq \left(-\infty, \tfrac{1}{2}\right)} \right)
		&\subseteq 
		g^{-1} \circ \kappa_\bot^{-1} \left( \Set{K \in \Kb(\R)}{K \subseteq \left(-\infty, \tfrac{1}{2}\right)}\right)\\
		&= g^{-1}\left((-\infty, \tfrac{1}{2})\right)\\
		&\subseteq V.
	\end{align*}
	This shows that $F$ witnesses $V$.
\end{proof}

Unlike in the case of uniform envelopability we have a very useful sufficient criterion available that asserts uniform $\mathcal{R}$-envelopability.
Its proof is essentially identical to that of \cite[Theorem 4.28]{NeumannPhD}.

\begin{thm}\label{Theorem: S-universal envelope of compactly majorisable function}
	Let $f \colon X \to Y$ be a function between computable $T_0$ spaces.
	Assume that $Y$ is effectively countably based and effectively Hausdorff.
	Assume that there exists a continuous map $B \colon X \to \K(Y)$ such that $f(x) \subseteq B(x)$.
	Then there exists a best continuous approximation $H\colon X \to \K(Y)$ of $\kappa \circ f$. 
	The induced $\O^2(Y)$-envelope $F(x) = \Set{U \in \O(Y)}{K \subseteq U}$ 
	is uniformly $\mathcal{R}$-universal.
\end{thm}

We obtain the result announced in the introduction:

\begin{prop}\label{Proposition: modulus of regularity}
    Let $f \colon X \to Y$ be a function sending a computable metric space $X$ to a computable $T_0$ space $Y$.
    Assume that $f$ admits an $\mathcal{R}$-universal envelope $F \colon X \to L$.
    Let $\widetilde{X}$ be a complete effective metric space and let $Z$ be an effective metric space.
    Let 
    $\beta \colon \widetilde{X} \times Y \to Z$ 
    and 
    $\alpha \colon \widetilde{X} \to X$ 
    be continuous functions.
	Fix an embedding $i \colon Z \to [0,1]^{\N}$.
    Then the following are equivalent:
    \begin{enumerate}
        \item There exists a relatively computable function 
        $
            \omega \colon \widetilde{X} \times \Q_+ \rightrightarrows \Q_+
        $
        such that 
        \begin{align}\label{eq: omega}
            &\forall \widetilde{x} \in \widetilde{X}.
            \forall \varepsilon > 0.
			\forall \delta \in \omega(\widetilde{x}, \varepsilon).
            \forall x \in X.\nonumber\\
            &\left(
                d(\alpha(\widetilde{x}), x) < \delta
                \rightarrow 
                d(\beta(\widetilde{x}, f \circ \alpha(\widetilde{x})), 
                    \beta(\widetilde{x}, f(x))) < \varepsilon
            \right).
		\end{align}
        \item 
		There exists a continuous function 
        $\widetilde{i \circ \beta} \colon \widetilde{X} \times L \to \K([0,1]^{\N})$
		satisfying 
		\[
			\widetilde{i \circ \beta}(\widetilde{x}, \xi_L(y)) \supseteq \{i \circ \beta(\widetilde{x}, y)\}
		\]
		for all $\widetilde{x} \in \widetilde{X}$ and all $y \in Y$, and
        \[
			\widetilde{i \circ \beta}(\widetilde{x}, F \circ \alpha(\widetilde{x})) = \{i \circ \beta(\widetilde{x}, f \circ \alpha(\widetilde{x}))\}
		\] 
		for all $\widetilde{x} \in \widetilde{X}$.
    \end{enumerate}
\end{prop}
\begin{proof}
	By Propositions 
	\ref{Proposition: Kb(Y) regular} 
	and 
	\ref{Proposition: regular approximation spaces closed under pullbacks}
	the approximation space $\kappa \circ i$ is regular.
	By Proposition \ref{Proposition: regular exponential}
	so is the approximation space 
	$(\kappa \circ i)_* \colon Z^{\widetilde{X}} \to \K([0,1]^{\N})^{\widetilde{X}}$.
	It then follows from the proof of
	\cite[Theorem 4.34]{NeumannPhD} that an extension of 
	$i \circ \beta$ 
	as in the second item exists 
	if and only if for every point $\widetilde{x} \in \widetilde{X}$ the point 
	$(\widetilde{x},\alpha(\widetilde{x})) \in \widetilde{X} \times X$
	is a point of continuity of the function
	\[
		\psi \colon \widetilde{X} \times X \to Z,
		\;
		\psi(\widetilde{x}, x) = \beta(\widetilde{x}, f(x)).
	\]

	Assume that there exists a relatively computable function 
	$\omega \colon \widetilde{X} \times \Q_+ \rightrightarrows \Q_+$ as above.
	We claim that every point 
	$(\widetilde{x}, \alpha(\widetilde{x})) \in \widetilde{X} \times X$ 
	is a point of continuity of the function 
	$\psi$.
	Let $\varepsilon > 0$.
	Feed $(\widetilde{x}, \varepsilon/3)$ into a relativised algorithm for computing a realiser of $\omega$.
	The algorithm will halt after finitely many steps and output a rational number $\delta > 0$ as in \eqref{eq: omega}.
	Since the algorithm halts within finitely many steps and $\widetilde{X}$ is countably based, 
	there exists an open neighbourhood $U$ of $\widetilde{x}$ such that for all $\widetilde{x}'$
	and all $x \in X$ we have the implication
	\begin{equation}\label{eq: modulus eq 1}
                d(\alpha(\widetilde{x}'), x) < \delta
                \rightarrow 
                d(\beta(\widetilde{x}', f \circ \alpha(\widetilde{x}')), 
                    \beta(\widetilde{x}', f(x))) < \varepsilon/3.
	\end{equation}
	Since $\alpha$ is continuous we may assume that $U$ is such that 
	$d(\alpha(\widetilde{x}), \alpha(\widetilde{x}')) < \delta/2$
	for all 
	$\widetilde{x}' \in U$.
	Since the function 
	$\widetilde{x}' \mapsto \beta(\widetilde{x}', f \circ \alpha(\widetilde{x}))$ 
	is continuous,
	we may further assume that $U$ is chosen such that we have
	$d(\beta(\widetilde{x}', f \circ \alpha(\widetilde{x})), \beta(\widetilde{x}, f \circ \alpha(\widetilde{x}))) < \varepsilon/3$
	for all $\widetilde{x}' \in U$. 

	Now, let $(\widetilde{x}', x) \in U \times B(\alpha(\widetilde{x}), \delta/2)$.
	We have 
	\begin{align*}
		d(\beta(\widetilde{x}', f(x)), \beta(\widetilde{x}, f \circ \alpha(\widetilde{x})))
		&\leq 
		d(\beta(\widetilde{x}', f(x)), \beta(\widetilde{x}', f \circ \alpha(\widetilde{x}')))\\
		&+d(\beta(\widetilde{x}', f \circ \alpha(\widetilde{x}')), \beta(\widetilde{x}', f \circ \alpha(\widetilde{x})))\\
		&+d(\beta(\widetilde{x}', f \circ \alpha(\widetilde{x})), 
		  \beta(\widetilde{x} , f \circ \alpha(\widetilde{x})))\\
		&< 
		\varepsilon.
	\end{align*}
	In the above, the first quantity in the sum is below $\varepsilon/3$ by a direct application of \eqref{eq: modulus eq 1}.
	The second quantity is below $\varepsilon/3$ by an application of \eqref{eq: modulus eq 1} with $x = \alpha(\widetilde{x})$.
	The third quantity is below $\varepsilon/3$ by the assumption on $U$.
	Thus, the preimage of the ball 
	$B(\beta(\widetilde{x},f \circ \alpha(\widetilde{x})), \varepsilon)$
	under the function $\psi$
	is a neighbourhood of $(\widetilde{x}, \alpha(\widetilde{x}))$.
	It follows that $(\widetilde{x}, \alpha(\widetilde{x}))$ is a point of continuity of $\psi$.

	Conversely, assume that every point of the form 
	$(\widetilde{x}, \alpha(\widetilde{x}))$ 
	is a point of continuity of the function 
	$\psi(x_0, x_1) = \beta(x_0, f(x_1))$.

	Fix a dense sequence $(z_n)_n$ in $Z$.
	For each ball $B(z_n, r)$ with rational radius $r$ there exists an 
	index $e_{z_n, r} \in \N$ 
	of a Turing machine and an oracle 
	$\Phi_{z_n, r} \colon \N \to \N$
	such that $e$ with oracle $\Phi$ computes the interior of 
	$\psi^{-1}(B(z_n, r))$.
	By standard coding arguments we can encode the data 
	$\langle z_n, r, e_{z_n, r}, \Phi_{z_n, r} \rangle$
	where $n$ ranges over $\N$ and $r$ ranges over $\Q_+$
	in a single sequence $(c_n)_n$ of positive integers such that for given 
	$n \in \N$, $r \in \N$, and $m \in \N$ the numbers $z_n$, $r$, $e_{z_n, r}$ and $\Phi_{z_n, r}(m)$
	are computable from $(c_n)_n$.

	We next describe an algorithm which computes a multi-valued function of type 
	$\widetilde{X} \times \Q_{+} \rightrightarrows \Q_{+}$ 
	relative to the sequence $(c_n)_n$ as an oracle.
	Given 
	$\widetilde{x} \in \widetilde{X}$ and $\Q \ni \varepsilon > 0$,
	apply the following partial algorithm to all elements of the sequence 
	$\langle z_n, r, e_{z_n, r}, \Phi_{z_n, r} \rangle$
	with $r \leq \varepsilon/2$ in parallel:
	Use $e_{z_n, r}$ with oracle $\Phi_{z_n, r}$ to test if 
	$(\widetilde{x}, \alpha(\widetilde{x})) \in \intr{\psi^{-1}(B(z_n, r))}$.
	If this test succeeds, then the algorithm $e_{z_n, r}$ with oracle $\Phi_{z_n, r}$
	has only read a finite initial segment of the given name of $\alpha(\widetilde{x})$.
	We can extract from this a $\delta > 0$ such that the algorithm halts for all 
	$x$ with $d(x, \alpha(\widetilde{x})) < \delta$.
	Now halt and output this $\delta$.

	We claim that this algorithm halts for all inputs $(\widetilde{x}, \varepsilon)$
	and outputs a $\delta$ as in \eqref{eq: omega}.
	
	If the algorithm halts, then the parallel search halts for some tuple 
	$\langle z_n, r, e_{z_n, r}, \Phi_{z_n, r} \rangle$
	with $r \leq \varepsilon/2$.
	The algorithm then outputs a $\delta > 0$ such that 
	$\psi(\widetilde{x}, x) \in B(z_n, r)$
	for all $x$ with $d(x, \alpha(\widetilde{x})) < \delta$.
	In particular, if $d(x, \alpha(\widetilde{x})) < \delta$ then 
	$d(\psi(\widetilde{x}, x), \psi(\widetilde{x}, \alpha(\widetilde{x}))) < \varepsilon$.

	It remains to show that the algorithm halts for all inputs $(\widetilde{x}, \varepsilon)$.
	Since the sequence $(z_n)_n$ is dense there exists a ball $B(z_n, r)$ with $r \leq \varepsilon/2$
	such that $\psi(\widetilde{x}, \alpha(\widetilde{x})) \in B(z_n, r)$.
	By assumption the point 
	$(\widetilde{x}, \alpha(\widetilde{x}))$ 
	is a point of continuity of $\psi$.
	It follows that 
	$(\widetilde{x}, \alpha(\widetilde{x}))$ 
	is contained in the interior of the preimage $\psi^{-1}(B(z_n, r))$.
	By construction the parallel search halts on the tuple 
	$\langle z_n, r, e_{z_n, r}, \Phi_{z_n, r} \rangle$.
\end{proof}

We obtain the following effective version of Proposition \ref{Proposition: modulus of regularity},
analogous to Corollary \ref{Corollary: uniform extension probes}:

\begin{cor}
    Let $f \colon X \to Y$ be a function which sends a computable metric space $X$ to a computable $T_0$ space $Y$.
    Assume that $f$ is uniformly $\mathcal{R}$-envelopable with uniform $\mathcal{R}$-envelope $F \colon X \to L$.
	Let $\widetilde{X}$ be a complete effective metric space and let $Z$ be an effective metric space.
    Let 
    $\beta \colon \widetilde{X} \times Y \to Z$ 
    and 
    $\alpha \colon \widetilde{X} \to X$ 
    be continuous functions.
	Fix an embedding $i \colon Z \to [0,1]^{\N}$.
	Let 
	\[ 
		h \colon Y \to \K\left([0,1]^{\N}\right)^{\widetilde{X}},
		\; 
		h(y)(\widetilde{x}) = \{ i \circ \beta(\widetilde{x}, y) \}.
	\]
    Then the following are equivalent:
    \begin{enumerate}
        \item There exists a relatively computable function 
        $
            \omega \colon \widetilde{X} \times \Q_+ \rightrightarrows \Q_+
        $
        such that 
        \begin{align*}
            &\forall \widetilde{x} \in \widetilde{X}.
            \forall \varepsilon > 0.
			\forall \delta \in \omega(\widetilde{x}, \varepsilon).
            \forall x \in X.\nonumber\\
            &\left(
                d(\alpha(\widetilde{x}), x) < \delta
                \rightarrow 
                d(\beta(\widetilde{x}, f \circ \alpha(\widetilde{x})), 
                    \beta(\widetilde{x}, f(x))) < \varepsilon
            \right).
		\end{align*}
        \item 
		We have 
		\[
			E_{L,\K([0,1]^{\N})^{\widetilde{X}}}(h)(F \circ \alpha (\widetilde{x}))(\widetilde{x})
			=
			\{i \circ \beta(\widetilde{x}, f \circ \alpha(\widetilde{x}))\}.
		\]
    \end{enumerate}
\end{cor}

\subsection{Composition of uniformly \texorpdfstring{$\mathcal{R}$}{R}-universal envelopes}

There is an analogous result to
Theorem \ref{Theorem: characterisation of universality for general envelopes} 
for $\mathcal{R}$-universal envelopes.
For an effective $T_0$ space $X$ let  
$
    \Sheaf{X}{-}
$
denote the sheaf of continuous real functions with values in $\R$.
An element of $\Sheaf{X}{-}$ is a continuous function $\varphi \colon V \to \R$
with $V \in \O(X)$.
In this context we will denote this function by $(V, \varphi)$.
We write $V = \dom(V,\varphi)$.
We write $\Sheaf{X}{X} \subseteq \Sheaf{X}{-}$ for the subset of \emph{global sections}, \textit{i.e.}, 
the set of total maps $\varphi \colon X \to \R$.
We will not need to endow the set $\Sheaf{X}{-}$ with a topology.

Any function 
$f \colon X \to Y$
induces a function
$
    \intr{f}_{\R} \colon \Sheaf{Y}{-} \to \Sheaf{X}{-}
$
which sends a section 
$(V, \varphi)$ 
to 
$(U, \varphi \circ f)$
where $U$ is the largest open set such that the restriction 
$\varphi \circ f|_U$
is well-defined and continuous.

We have for all $V \in \O(Y)$ a map 
$
	i_V \colon \O(V) \to \O(Y),
$
which sends $U \in \O(V)$ to $U \in \O(Y)$.
This induces a map 
$
	i_V^* \colon \O^2(Y) \to \O^2(V)
$.
A continuous map 
$F \colon X \to \O^2(Y)$
thus induces the map 
$
    \starred{F}_{\R} \colon \Sheaf{Y}{-} \to \Sheaf{X}{-}
$
which sends 
$(V, \varphi)$
to 
$(U, \nu_{\R}^{-1} \circ \varphi^{**} \circ i_V^* \circ F)$,
where 
$U$
is the largest open set such that 
$\varphi^{**} \circ i_V^*  \circ F(x) \in \nu_{\R}(\R)$
for all $x \in U$.

It is easy to see that if $F$ is an envelope of $f$ then for all 
$(V, \varphi) \in \Sheaf{Y}{-}$
we have 
$\dom \starred{F}_{\R}(V,\varphi) \subseteq \dom \intr{f}_{\R}(V,\varphi)$.

\begin{thm}\label{Theorem: characterisation uniform R-universality}
    Let 
    $f \colon X \to Y$ be a function between effective $T_0$ spaces.
    Assume that $Y$ is a normal Hausdorff space.
    Let $F \colon X \to \O^2(Y)$ be the principal $\O^2(Y)$-envelope of $f$.
    Then $F$ is uniformly $\mathcal{R}$-universal if and only if 
    $\left(\starred{F}_{\R}\right)|_{\Sheaf{Y}{Y}} = \left(\intr{f}_{\R}\right)|_{\Sheaf{Y}{Y}}$.
\end{thm}
\begin{proof}
Assume that $F$ is uniformly $\mathcal{R}$-universal.
Let $\varphi \colon Y \to \R$ be a continuous function.
Let $U$ be the largest open set such that $\varphi \circ f$ is continuous on $U$.
Consider the envelope 
$G \colon X \to \O^2(\R)$
defined as 
$G(x) = \nu_{\R} \circ \varphi \circ f(x)$ if $x \in U$,
and 
$G(x) = \emptyset$ otherwise,
with inclusion map 
$\nu_{\R} \circ \varphi \colon Y \to \O^2(\R)$.
Then by Propositions 
\ref{Proposition: Kb(Y) regular} 
and 
\ref{Proposition: regular approximation spaces closed under pullbacks}
the envelope $G$ is a regular envelope of $f$.
Since $F$ is assumed to be uniformly $\mathcal{R}$-universal we have  
$\varphi^{**} \circ F(x) = G(x)$
for all $x \in U$.
The claim follows.
    
Now assume that $\starred{F}_{\R} = \intr{f}_{\R}$ on global sections of $\Sheaf{Y}{-}$.
Let $G \colon X \to M$ be a regular envelope of $f$.
Let $G(x) \in V \in \O(M)$.
Then there exists $U \in \O(M)$ with $G(x) \in U$ and 
\[
    \xi_M^{-1}(\ucl{G(x)}) \subseteq \xi_M^{-1}(U) \subseteq \clos{\xi_M^{-1}(U)} \subseteq \xi_M^{-1}(V).
\] 
By Urysohn's lemma there exists a continuous function 
$\varphi \colon Y \to \R$
with $\varphi(y) = 1$ for all $y \in \clos{\xi_M^{-1}(U)}$
and $\varphi(y) = 0$ for all $y \in Y \setminus \xi_M^{-1}(V)$.

The set $\xi_M^{-1}(U)$ is a robust property of $f(x)$ so that there exists a neighbourhood $W$ of $x$ that gets mapped to $\xi_M^{-1}(U)$ by $f$.
It follows that $\varphi \circ f$ is constant equal to $1$ on $W$.
By assumption $\varphi^{**} \circ F$ is constant equal to $\nu_{\R}(1)$ on $W$.
In particular $\varphi^{*}\left((+\tfrac{1}{2},+\infty)\right) \in F(x)$. 
Now,
$\varphi^{*}\left((\tfrac{1}{2},+\infty)\right) \subseteq \xi_M^*(V)$,
so that 
$\xi_M^*(V) \in F(x)$.
Since $V$ was an arbitrary set with $G(x) \in V$ it follows that $F$ uniformly tightens $G$.
\end{proof}

The restriction to global sections of $\Sheaf{Y}{-}$ in Theorem \ref{Theorem: characterisation uniform R-universality}
is essential.
Observe that continuous functions of type 
$\varphi \colon Y \to \Sigma$
can be modelled by partial continuous functions 
$\varphi \colon V \to \R$
with open domain.
It follows that if
$\starred{F}_\R = \intr{f}_\R$
on all of 
$\Sheaf{Y}{-}$
then 
$\starred{F} = \intr{f}$,
so that $F$ is uniformly universal (for the class of all envelopes).

We will now prove a composition theorem analogous to Theorem \ref{Theorem: characterisation of universality for general envelopes}.
The following technical result is easily verified:

\begin{lem}\label{Lemma: restriction and i_U}
	Let $X$ and $Y$ be effective $T_0$ spaces.
	Let $U \in \O(X)$ be an open subset of $X$.
	Let 
	$\varphi \colon X \to Y$ 
	and 
	$\psi \colon U \to Y$ 
	be continuous functions such that $\varphi|_U = \psi$.
	Then $\varphi^* \geq i_U \circ \psi^*$.
\end{lem}

\begin{lem}
    Let 
    $F \colon X \to \O^2(Y)$ 
    and 
    $G \colon Y \to \O^2(Z)$
	be continuous functions.
	Then we have 
    \[
        \dom \left(\starred{(G \bullet F)}_{\R}(V, \varphi)\right) \supseteq \dom \left(\starred{F}_{\R} \circ \starred{G}_{\R} (V,\varphi)\right)
    \]
	for all $(V, \varphi) \in \Sheaf{Z}{-}$.
\end{lem}
\begin{proof}
    By definition we have 
	\begin{equation}\label{eq: R-composition eq 1}
		\starred{(G \bullet F)}_{\R}(V,\varphi) =
		(U, \nu_{\R}^{-1} \circ \varphi^{**} \circ i_V^* \circ \nu_{\O(Z)}^* \circ G^{**} \circ F),
	\end{equation}
	\[
		\starred{G}_{\R} (V, \varphi)
		=
		(T, \nu_{\R}^{-1} \circ \varphi^{**} \circ i_V^* \circ G|_T),
	\]
	and 
	\[
		\starred{F}_{\R} \circ \starred{G}_{\R} (V, \varphi)
		= 
		(W, \nu_{\R}^{-1} \circ (\nu_{\R}^{-1})^{**} \circ \varphi^{****} \circ i_V^{***} \circ G|_T^{**} \circ i_T^* \circ F).
	\]
	The unit and multiplication laws for the monad $\O^2$ imply
	\[
		\varphi^{**} \circ i_V^* \circ \nu_{\O(Z)}^*
		=
		\nu_{\O(\R)}^* \circ \varphi^{****} \circ i_V^{***}.
	\]
	Plugging this into \eqref{eq: R-composition eq 1} we obtain:
	\[
		\starred{(G \bullet F)}_{\R}(V,\varphi) =
		(U, \nu_{\R}^{-1} \circ \nu_{\O(\R)}^* \circ \varphi^{****} \circ i_V^{***} \circ G^{**} \circ F)
	\]
	On the open set $T \in \O(Y)$ we have
	$\varphi^{**} \circ i_V^* \circ G|_T = \nu_{\R} \circ \Phi$
	for some function 
	$\Phi \colon T \to \R$.
	Using Lemma \ref{Lemma: restriction and i_U} we obtain:
	\[ 
	\left(\varphi^{**} \circ i_V^* \circ G\right)^{**}
	\geq 
	(\nu_{\R} \circ \Phi)^{**} \circ i_T^*
	=
	\left(\varphi^{**} \circ i_V^* \circ G|_T\right)^{**} \circ i_T^*
	\]
	We hence obtain:
	\begin{align*}
		\nu_{\O(\R)}^{*} \circ \varphi^{****} \circ i_V^{***} \circ G^{**} \circ F
		&\geq 
		\nu_{\O(\R)}^* \circ \nu_{\R}^{**} \circ \Phi^{**} \circ i_T^* \circ F\\
		&=
		\Phi^{**} \circ i_T^* \circ F\\
		&=
		(\nu_{\R}^{-1})^{**} \circ \varphi^{****} \circ i_V^{***} \circ G|_T^{****} \circ i_T^* \circ F.
	\end{align*}
	This implies by the definition of $W$ and $U$ that $W \subseteq U$.
\end{proof}

We obtain a characterisation analogous to Theorem \ref{Theorem: characterisation of universality for general envelopes}
and a number of corollaries.

\begin{thm}
    Let 
    $f \colon X \to Y$ 
    and 
    $g \colon Y \to Z$
    be functions between effective $T_0$ spaces.
	Assume that $Y$ and $Z$ are normal Hausdorff spaces.
    Let 
    $F \colon X \to \O^2(Y)$
    and 
    $G \colon Y \to \O^2(Z)$
    be envelopes of 
    $f$ and $g$
    respectively.
	Assume that $F$ and $G$ are uniformly $\mathcal{R}$-universal.
	Then 
	$(G \bullet F)$
	is a uniformly $\mathcal{R}$-universal envelope of 
	$g \circ f$ 
	if and only if 
	\[ 
		\left(\intr{(g \circ f)}_{\R}\right)|_{\Sheaf{Z}{Z}}
		=
		\left(\intr{f}_{\R} \circ \intr{g}_{\R}\right)|_{\Sheaf{Z}{Z}}.
	\] 
\end{thm}

\begin{cor}
	Let 
    $f \colon X \to Y$ 
    and 
    $g \colon Y \to Z$
    be functions between effective $T_0$ spaces.
    Assume that $Z$ is a normal Hausdorff space.
    Let 
    $F \colon X \to \O^2(Y)$
    and 
    $G \colon Y \to \O^2(Z)$
    be envelopes of 
    $f$ and $g$
    respectively.
	Assume that $F$ is uniformly universal and that 
	$G$ is uniformly $\mathcal{R}$-universal.
	Then 
	$G \bullet F$
	is a uniformly $\mathcal{R}$-universal envelope of 
	$g \circ f$ 
	if and only if 
	\[
		\intr{(g \circ f)}_{\R}
		=
		\intr{f}_{\R} \circ \intr{g}_{\R}.
	\]
\end{cor}

\begin{cor}
	Let 
    $f \colon X \to Y$ 
    and 
    $g \colon Y \to Z$
    be functions between effective $T_0$ spaces.
    Assume that $Z$ is a normal Hausdorff space.
    Let 
    $F \colon X \to \O^2(Y)$
    and 
    $G \colon Y \to \O^2(Z)$
    be envelopes of 
    $f$ and $g$
    respectively.
	Assume that $F$ is uniformly universal and that 
	$G$ is uniformly $\mathcal{R}$-universal.
	If $f$ is open or $g$ is continuous then 
	$G \bullet F$
	is a uniformly $\mathcal{R}$-universal envelope of 
	$g \circ f$.
\end{cor}

The assumption that $F$ be uniformly universal rather than just $\mathcal{R}$-universal
cannot be weakened in general, as the next proposition shows.

\begin{prop}
	Let 
    $f \colon X \to Y$ 
	be a function between effective $T_0$ spaces.
	Let $F \colon X \to \O^2(Y)$ be its principal $\O^2(Y)$-envelope.
	Let $Z$ be a metric space which contains at least two points.
	Assume that for all uniformly $\mathcal{R}$-envelopable functions 
	$g \colon Y \to Z$ 
	with principal envelope 
	$G \colon Y \to \O^2(Z)$
	the composition 
	$G \bullet F$
	is uniformly $\mathcal{R}$-universal.
	Then $F$ is uniformly universal.
\end{prop}
\begin{proof}
	Let $V \in \O(Y)$ be a robust property of $f(x)$.
	Fix two points $a, b \in Z$
	with $a \neq b$.
	Let 
	$g \colon Y \to Z$
	be the function that maps 
	$V$ to $t$ 
	and $Y \setminus V$ to $b$.
	Then by Theorem \ref{Theorem: S-universal envelope of compactly majorisable function}
	$g$ is uniformly 
	$\mathcal{R}$-envelopable
	with principal envelope
	\[ 
		G(x) =  
		\begin{cases}
			\nu_Z(a) &\text{if }x \in V, \\ 
			\nu_Z(b) &\text{if }x \in \intr(Y \setminus V), \\
			\nu_Z(a) \cap \nu_Y(b) &\text{otherwise.}
		\end{cases}
	\]	
	By assumption the composition
	$G \bullet F(x)$
	is uniformly $\mathcal{R}$-universal.
	In particular it is equal to the principal $\O^2(Y)$-envelope of $g \circ f$.
	Now, since $V$ is a robust property of $f(x)$ there exists $U \in \O(X)$ such that 
	$x \in U \subseteq f^{-1}(V)$.
	It follows that 
	\[
		H \colon X \to \O^2(Z),
		\; 
		H(x) = 
		\begin{cases}
			\nu_Z(a) &\text{if }x \in U \\ 
			\emptyset &\text{otherwise.}
		\end{cases}
	\]
	is an envelope of $g \circ f$.
	We thus obtain $G \bullet F(x) = \nu_Y(a)$.
	Let $W \in \O(Z)$ be an open set with $a \in W$ and $b \notin W$.
	Then by Lemma \ref{Lemma: star respects composition} we have
	$x \in \starred{(G \bullet F)}(W) = \starred{F}(V)$,
	so that $F$ is uniformly universal by Theorem \ref{Theorem: motivation}.
\end{proof}

\section{Advice bundles and co-envelopes}

We can assign to any function $f \colon X \to Y$ a canonical envelope that generalises the principal $\O^2(Y)$-envelope of uniformly envelopable functions.
On these envelopes we can define a -- in general non-associative -- composition operation which generalises composition of uniform envelopes.
In order to achieve this we need to first pass to a dual notion.
The point of departure is the observation that we can identify a uniform envelope $F \colon X \to \O^2(Y)$ with the map 
$\starred{F} \colon \O(Y) \to \O(X)$ which approximates the function $\intr{f} \colon \O(Y) \to \O(X)$.

Let $Y$ be an effective $T_0$ space.
An \emph{advice bundle}
$(A,\pi_A)$
over $Y$ is an injective space $A$ 
together with a continuous map 
$
    \pi_A \colon A \to \O(Y)
$ 
which preserves finite meets and arbitrary joins,
and which admits a continuous section
$s \colon \O(Y) \to A$.

Let $f \colon X \to Y$ be a function.
A \emph{co-envelope} of $f$ is an advice bundle $A$ over $\O(Y)$ together with a continuous map
$\starred{F} \colon A \to \O(X)$
satisfying 
$\starred{F} \leq \intr{f} \circ \pi_A$.
As in the case of envelopes, for all advice bundles $A$ over $\O(Y)$ there exists a greatest continuous 
map $\starred{F} \colon A \to \O(X)$ satisfying $\starred{F} \leq \intr{f} \circ \pi_A$.
We call this the \emph{principal $A$-co-envelope}.

If 
$\starred{F} \colon A \to \O(X)$
and 
$\starred{G} \colon B \to \O(X)$
are co-envelopes of $f$ then 
$\starred{F}$ 
\emph{tightens} 
$\starred{G}$ 
if there exists a continuous map 
$\Phi \colon B \to A$
such that 
$\pi_A \circ \Phi = \pi_B$
and 
$\starred{F} \circ \Phi \geq \starred{G}$.
A co-envelope of $f$ is called \emph{universal} if it tightens all co-envelopes of $f$.

The name ``advice bundle'' is motivated as follows:
Since the map $\pi_A$ preserves joins and meets, the fibre
$\pi_A^{*}(V)$ 
of an open set $V \in \O(Y)$ under $\pi_A$ is a lattice.
By virtue of a section $s \colon \O(Y) \to A$
the set $V$ can be identified with the element $s(V) \in \pi_A^{*}(V)$.
The elements of $\pi_A^{*}(V)$ above $s(V)$ can be interpreted as 
``$V$ enriched with additional information''.

Envelopes and co-envelopes are connected as follows:
For a function 
$f \colon X \to Y$ 
let 
$\mathcal{E}(f)$ 
denote the category whose objects are envelopes 
$(F,\xi_L)$ of $f$ 
such that $\xi_L^*$ has a continuous section 
$s \colon \O(Y) \to \O(L)$,
and whose morphisms 
$\Phi \colon (F, \xi_L) \to (G, \xi_M)$
are continuous maps 
$\Phi \colon L \to M$
satisfying 
$\Phi \circ \xi_L = \xi_M$
and 
$\Phi \circ F \geq G$.

It is easy to see that the pre-order one obtains from $\mathcal{E}(f)$ by collapsing all morphisms into a single arrow 
is isomorphic to the pre-order of envelopes of $f$ with the tightening relation:
An envelope 
$F \colon X \to L$ 
with inclusion map 
$\xi_L \colon Y \to L$
can be made into the envelope
$H \times F \colon X \to \O^2(Y) \times L$,
where $H$ is the constant function with value $\emptyset$,
with inclusion map 
$\nu_Y \times \xi_L$.
If $F$ tightens $G$ 
then $H \times F$ tightens $H \times G$ 
via the greatest continuous extension 
$(\nu_Y \times \xi_M)/(\nu_Y \times \xi_L)$
of $(\nu_Y \times \xi_M)$ along $(\nu_Y \times \xi_L)$,
which is easily seen to satisfy 
$(\nu_Y \times \xi_M)/(\nu_Y \times \xi_L) \circ (\nu_Y \circ \xi_L) = (\nu_Y \times \xi_M)$.
Thus, for our purpose, the additional restrictions put on envelopes and the tightening relation are inessential.

Let 
$\mathcal{A}(f)$
denote the category whose objects are co-envelopes
$(\starred{F}, \pi_A)$ of $f$
and whose morphisms
$\Phi \colon (\starred{F}, \pi_A) \to (\starred{G}, \pi_B)$
are continuous maps 
$\Phi \colon B \to A$
that witness the tightening relation.
Then we have a functor 
$\mathcal{F} \colon \mathcal{A}(f) \to \mathcal{E}(f)$
that sends objects
$(\starred{F}, \pi_A)$
to 
$((\starred{F})^* \circ \nu_X, \pi_A^* \circ \nu_Y)$
and morphisms
$\Phi \colon (\starred{F}, \pi_A) \to (\starred{G}, \pi_B)$
to 
$\Phi^* \colon \O(A) \to \O(B)$.
In the other direction we have a functor 
$\mathcal{G} \colon \mathcal{E}(f) \to \mathcal{A}(f)$
that sends objects
$(F, \xi_L)$
to 
$(F^*, \xi_L^*)$
and morphisms
$\Phi \colon L \to M$
to 
$\Phi^* \colon \O(M) \to \O(L)$.
One easily verifies that $\mathcal{F}$ is a left adjoint of $\mathcal{G}$.
Any co-envelope
$(\starred{F}, \pi_A)$
is naturally tightened by 
$\mathcal{G} \circ \mathcal{F}(\starred{F}, \pi_A)$
and any envelope 
$(F, \xi_L)$
naturally tightens 
$\mathcal{F} \circ \mathcal{G}(F,\xi_L)$.
Conversely, $\mathcal{F} \circ \mathcal{G}(F,\xi_L)$ tightens $(F, \xi_L)$,
and $\mathcal{G} \circ \mathcal{F} (\starred{F}, \pi_A)$ tightens $(\starred{F},\pi_A)$,
but this tightening is not natural.
Nonetheless, if we forget about the specific choice of map for the tightening and thus make 
$\mathcal{E}(f)$ and $\mathcal{A}(f)$ into pre-orders, 
then the functors $\mathcal{F}$ and $\mathcal{G}$ induce an equivalence. 

In particular every function has a universal co-envelope, which corresponds to a universal envelope.

We have the following straightforward dual to Scott's observation \cite{ScottLattices}. 

\begin{prop}\label{Proposition: greatest extension dual}
    Let 
    $A, B, C$ 
    be effective complete lattices.
    Let 
    $\sigma \colon B \to C$ 
    and 
    $\rho \colon C \to B$
    be continuous maps 
    with 
    $\rho \circ \sigma = \id_B$
    such that 
    $\rho$ 
    preserves arbitrary joins.
    Let 
    $\varphi \colon A \to B$.
    Then there exists a greatest continuous 
    map 
    $\lift{\rho}{\varphi} \colon A \to C$
    satisfying 
    $
        \rho \circ \lift{\rho}{\varphi} = \varphi
    $.
\end{prop}

An advice bundle
$\pi_E \colon E \to \O(Y)$ 
is called a \emph{least advice bundle for $f \colon X \to Y$} if for all advice bundles 
$\pi_A \colon A \to \O(Y)$ 
such that the principal $A$-co-envelope of $f$ is universal,
there exists a unique continuous map 
$r \colon A \to E$
which admits a continuous section 
$s \colon E \to A$,
such that 
$r$ 
witnesses the tightening of the principal $A$-envelope $\starred{P}_A$
by the principal $E$-envelope $\starred{P}_E$
and $s$ witnesses the tightening in the other direction.
It follows immediately from the definition that the principal co-envelope associated with a least advice bundle is universal.

Note that a least advice bundle, should it exist, is unique up to unique isomorphism.
If $f$ is uniformly envelopable then $\O(Y)$ is the least advice bundle for $f$ and 
the principal $\O(Y)$-co-envelope can be identified with the principal $\O^2(Y)$-envelope.

\begin{thm}\label{Theorem: least complete advice bundle}
    Let 
    $f \colon X \to Y$ 
    be a function between effective $T_0$ spaces.
    Then $f$ has a least advice bundle 
    $\mathfrak{A}_f$.
\end{thm}
\begin{proof}
    Let 
    \[
        L_f = \Set{(U,V) \in \O(X) \times \O(Y)}
                  {U \subseteq f^{-1}(V)}.
    \]
    Let $\pi_X \colon L_f \to \O(X)$ be the map that sends $(U,V)$ to $U$.
    Let $\pi_Y \colon L_f \to \O(Y)$ be the map that sends $(U,V)$ to $V$. 
    Observe that the space $L_f$ is an effective complete lattice:
    $\O(X)$ and $\O(Y)$ are effective complete lattices, so that 
    the product $\O(X) \times \O(Y)$ is an effective complete lattice by 
    \cite[Proposition 3.5]{NeumannPhD}.
    Further, by the same proposition, overt joins and compact meets in $\O(X) \times \O(Y)$ are given
    by component-wise union and intersection respectively.
    Since the map $f^{-1} \colon 2^{\O(Y)} \to 2^{\O(X)}$ preserves arbitrary unions and intersections,
    it follows that $L_f \subseteq \O(X) \times \O(Y)$ is closed under overt joins and compact meets.
    Now, \cite[Proposition 3.4]{NeumannPhD} yields that $L_f$ is an effective complete lattice, as claimed.
    By Proposition \ref{Proposition: greatest extension dual} there exists a greatest continuous map 
    $P_f \colon L_f \to L_f$
    with 
    $\pi_Y \circ P_f = \pi_Y$.
    Since 
    $\pi_Y \circ \id_{L_f} = \pi_Y$ 
    and 
    $\pi_Y \circ P_f^2 = \pi_Y$
    we have 
    $P_f^2 = P_f$.
    It follows that $P_f$ is a projection onto its range $\mathfrak{A}_f$.
    It follows from \cite[Proposition 3.7]{NeumannPhD} that $\mathfrak{A}_f$ is an effective complete lattice.
    The map 
    $\pi_Y \colon \mathfrak{A}_f \to \O(Y)$
    preserves arbitrary joins and finite meets.
    It has the continuous section 
    \[
        \sigma \colon \O(Y) \to \mathfrak{A}_f,
        \;
        \sigma(V) = P_f(\emptyset, V).
    \]
    We claim that
    $\mathfrak{A}_f$ 
    is a least advice bundle.
    We will defer the proof that $\mathfrak{A}_f$ is injective until the very end.
    Nonetheless we will apply the terminology we have introduced above for advice bundles 
    to $\mathfrak{A}_f$ already.
    
    We claim that the principal $\mathfrak{A}_f$-co-envelope of $\intr{f}$ is given by $\pi_X$.
    Let 
    $\starred{F} \colon \mathfrak{A}_f \to \O(X)$
    be a co-envelope of 
    $\intr{f}$.
    Then 
    $\starred{F}(U,V) \subseteq \intr{f}(V)$.
    It follows that the map 
    \[
        L_f \to L_f,
        \;
        (U,V) \mapsto (\starred{F} \circ P_f(U,V), V)
    \]
    is well-defined.
    It is hence below $P_f$.
    But this means 
    $\starred{F}(U,V) \subseteq U$ 
    for all 
    $(U,V) \in \mathfrak{A}_f$.
    Thus, $\pi_X$ is indeed the principal co-envelope.

    Let 
    $\pi_E \colon E \to \O(Y)$
    be an advice bundle such that the principal co-envelope 
    $\starred{G} \colon E \to \O(X)$
    is co-universal.

    Consider the map
    \[ 
        \Phi \colon E \to \mathfrak{A}_f,
        \;
        \Phi(x) = P_f(\starred{G}(x), \pi_E(x)).
    \]
    This map is well-defined, for 
    $\starred{G} \subseteq \intr{f} \circ \pi_E$
    since 
    $\starred{G}$ 
    is a co-envelope.
    It obviously satisfies $\pi_Y \circ \Phi = \pi_E$.
    Now, 
    $\pi_X \circ \Phi$ 
    with projection 
    $\pi_Y \circ \Phi = \pi_E$
    is an $E$-co-envelope of $f$.
    Since $\starred{G}$ is assumed to be the principal $E$-co-envelope 
    it follows that $\pi_X \circ \Phi = \starred{G}$.
    From this it easily follows that $\Phi$ is the only map that witnesses 
    the tightening of $\starred{G}$ by $\starred{F}$.

    Since $\starred{G}$ is assumed to be universal there exists a map
    $\Psi \colon \mathfrak{A}_f \to E$
    such that 
    $\pi_E \circ \Psi = \pi_{Y}$
    and 
    $\starred{G} \circ \Psi = \pi_X$.  
    
    Let $(U,V) \in \mathfrak{A}_f$.
    Then 
    \[ 
        \Phi \circ \Psi ((U,V))
        =
        P_f(\starred{G} \circ \Psi (U, V), \pi_E \circ \Psi (U,V))
        = 
        P_f(U, V)
        = 
        (U,V).
    \] 
    It follows that 
    $\mathfrak{A}_f$
    is a retract of 
    $E$
    as claimed.

    To make 
    $\mathfrak{A}_f$ 
    into a least advice bundle it remains to show that 
    $\mathfrak{A}_f$ 
    is an injective space.
    But using the above proof and the functors $\mathcal{F}$ and $\mathcal{G}$ defined above,
    we obtain that 
    $\O^2(\mathfrak{A}_f)$ 
    is a universal advice bundle,
    so that $\mathfrak{A}_f$
    is a retract of $\O^2(\mathfrak{A}_f)$ in a unique way.
    This shows that $\mathfrak{A}_f$ is indeed injective.
\end{proof}

Theorem \ref{Theorem: least complete advice bundle} yields an explicit construction 
of the least advice bundle $\mathfrak{A}_f$ of a given function $f \colon X \to Y$.
However, the specific representation of $\mathfrak{A}_f$ obtained in this construction is 
not very useful from a computational point of view, 
since the principal co-envelope is simply a projection.
The computational content of this envelope seems to be, so to speak, hidden in the complexity of the 
identification of $\mathfrak{A}_f$ with a subspace of $\O(X) \times \O(Y)$.

Let $f \colon X \to Y$ be a function between effective $T_0$ spaces.
Let $\mathfrak{A}_f$ be its least advice bundle.
We claim that the fibres $\pi_Y^*(V)$ are injective spaces.
For every 
$V \in \O(Y)$ 
the set 
$\pi_Y^*(V)$
is a retract of the lattice 
$L_f(V) = \Set{U \in \O(X)}{U \subseteq f^{-1}(V)}$.
The lattice $L_f(V)$ itself is a retract of $\O(X)$ under the continuous map 
$U \mapsto U \cap \intr{f^{-1}(V)}$.
Since injective spaces are closed under retracts, it follows that 
$\pi_Y^*(V)$
is injective.
We will also write $\mathfrak{A}_f(V)$ for $\pi_Y^*(V)$.

It is easy to see that the map 
$\sigma \colon \O(Y) \to \mathfrak{A}_f$
which sends 
$V \in \O(Y)$
to 
$P_f(\emptyset, V) \in \mathfrak{A}_f$
is a lower adjoint of 
$\pi_Y \colon \mathfrak{A}_f \to \O(Y)$.

The 
\emph{primary co-envelope} 
of 
$f \colon X \to Y$
is the principal co-envelope 
\[
    \starred{\E_f} \colon \mathfrak{A}_f \to \O(X).
\]
The 
\emph{primary envelope}
of 
$f$ 
is its dual 
\[
    \E_f \colon X \to \O(\mathfrak{A}_f).
\]

Let $f \colon X \to Y$ and $g \colon Y \to Z$ be functions.
Let 
$\starred{\E_g} \colon \mathfrak{A}_g \colon \O(Y)$
and
$\starred{\E_f} \colon \mathfrak{A}_f \to \O(X)$
be their respective primary co-envelopes.
Then the 
\emph{composition}
$\starred{\E_f} \bullet \starred{\E_g}$
is defined as:
\[
    \starred{\E_f} \bullet \starred{\E_g}
    =
    \starred{\E_f} \circ \lift{\pi_Y}{\starred{\E_g}}.
\]
This generalises the composition of principal $\O^2(Y)$-envelopes.
Due to the presence of the greatest continuous lift, the composition is in general no longer associative. 

We obtain an analogous result to
Theorem \ref{Theorem: characterisation of universality for general envelopes}:

\begin{thm}\label{Theorem: general composition theorem}
    Let $f \colon X \to Y$, 
        $g \colon Y \to Z$.
    Let 
    $\starred{\E_f} \colon \mathfrak{A}_f \to \O(X)$
    and 
    $\starred{\E_g} \colon \mathfrak{A}_g \to \O(Y)$
    be the primary co-envelopes of $f$ and $g$ respectively.
    Assume that either of the following two conditions is satisfied:
    \begin{enumerate}
        \item $f$ is open 
        \item $g$ is uniformly envelopable and $\intr{(g \circ f)} = \intr{f} \circ \intr{g}$.
    \end{enumerate}
    Then
    $\starred{\E_f} \bullet \starred{\E_g}$
    is the principal 
    $\mathfrak{A}_g$-co-envelope 
    of $g \circ f$.
\end{thm}
\begin{proof}
    Let 
    $\starred{H} \colon \mathfrak{A}_g \to \O(X)$ 
    be a $\mathfrak{A}_g$-co-envelope of 
    $g \circ f$.
    We use the representation of $\mathfrak{A}_g$ from Theorem \ref{Theorem: least complete advice bundle}.
    Let 
    $(U, V) \in \mathfrak{A}_g$.
    Then, under either of the above assumptions we have by Theorem \ref{Theorem: interior inverse and openness}:
    \begin{equation}\label{eq: composition thm eq}
        \starred{H}(U,V) 
        \subseteq 
        \intr{(g \circ f)}(V)
        = \intr{f} \circ \intr{g}(V).
    \end{equation}

    Assume that $f$ is open.
    Then,  
    since the spaces $\O(X)$ and $\O(Y)$ carry the Scott topology,
    the map $f_* \colon \O(X) \to \O(Y)$
    which sends $U \in \O(X)$ to $f(U) \in \O(Y)$
    is well-defined and continuous.
    Now \eqref{eq: composition thm eq}
    implies 
    $f_*(\starred{H}(U,V)) \subseteq \intr{g}(V)$,
    for all $(U, V) \in \mathfrak{A}_g$,
    so that the map 
    \[
        L_g \to L_g,
        \;
        (U, V) \mapsto (f_* \circ \starred{H} \circ P_g (U,V), V) 
    \]
    is well-defined and continuous.
    Here, $L_g$ and $P_g$ are defined as in the proof of Theorem \ref{Theorem: least complete advice bundle}.
    For $(U,V) \in \mathfrak{A}_g$ it follows from the definition of $P_g$ that  
    \[ 
        f_* \circ \starred{H} \circ P_g(U,V) 
        \leq 
        \starred{\E_g} \circ P_g(U,V)
        =
        \starred{\E_g} (U,V) 
        =
        \pi_Y(U,V)
        = U.
    \]
    This in turn implies that the map 
    \[
        \Phi \colon 
        \mathfrak{A}_g \to \mathfrak{A}_f,
        \; 
        (U,V) \mapsto P_f(\starred{H}(U,V), U)
    \]
    is well-defined and continuous.
    By definition of 
    $\lift{\pi_Y}{\starred{\E_g}}$
    we have 
    $\lift{\pi_Y}{\starred{\E_g}} \geq \Phi$.
    Hence 
    \[ 
        \starred{\E_f} \circ \lift{\pi_Y}{\starred{\E_g}}(U,V) 
        \geq \pi_X \circ \Phi(U,V) 
        \geq \starred{H}(U,V).
    \]
    This shows that 
    $\starred{\E_f} \circ \lift{\pi_Y}{\starred{\E_g}}$
    is greater than or equal to every co-envelope of $g \circ f$
    with values in $\mathfrak{A}_g$.
    It is therefore the principal $\mathfrak{A}_g$-co-envelope.

    Now assume that $g$ is uniformly envelopable.
    Then
    $\mathfrak{A}_g \simeq \O(Z)$
    and 
    $\starred{\E_g} = \intr{g} \colon \O(Z) \to \O(Y)$.
    Under this identification the map $\starred{H}$ becomes a map of type  
    $\starred{H} \colon \O(Z) \to \O(X)$ 
    with 
    $\starred{H}(V) \subseteq \intr{f}(\intr{g}(V))$.
    We obtain a well-defined continuous map 
    \[
        \O(Z) \to \mathfrak{A}_f,
        \;
        V \mapsto P_f(\starred{H}(V), \intr{g}(V)).
    \]
    By definition this map is smaller than or equal to $\lift{\pi_Y}{\intr{g}}$.
    It follows that $\starred{H}(V) \subseteq \pi_X \circ \lift{\pi_Y}{\intr{g}}(V)$.
    This proves the claim.
\end{proof}

Compared with 
Theorem \ref{Theorem: general composition theorem},
Theorem \ref{Theorem: characterisation of universality for general envelopes} has a stronger conclusion.
The composition is even asserted to be uniformly universal,
rather than just being the principal $\O^2(Z)$-envelope.
The next proposition constitutes somewhat of a converse to this.

\begin{prop}\hfill 
    \begin{enumerate}
    \item 
    Let $f \colon X \to Y$.
    Let $\starred{\E_f} \colon \mathfrak{A}_f \to \O(X)$ be its primary co-envelope.
    If for all effective $T_0$ spaces and all continuous functions 
    $g \colon Y \to Z$
    the composition 
    $\starred{\E_f} \bullet g^*$
    is uniformly universal then $f$ is uniformly envelopable.
    \item 
    Let $g \colon Y \to Z$.
    Let $\starred{\E_g} \colon \mathfrak{A}_g \to \O(X)$ be its primary co-envelope.
    If for all effective $T_0$ spaces and all continuous functions 
    $f \colon X \to Y$
    the composition 
    $f^* \bullet \starred{\E_g}$
    is uniformly universal then $g$ is uniformly envelopable.    
\end{enumerate}
\end{prop}
\begin{proof}
    For the first claim, take $g = \id_Y \colon Y \to Y$.
    Then $\starred{\E_f} \bullet g^* = \starred{\E_f}$, 
    so that $\mathfrak{A}_f \simeq \O(Y)$
    and $f$ is uniformly envelopable.
    For the second claim, take $f = \id_Y \colon Y \to Y$.
    Then 
    $f^* \bullet \starred{\E_g} = f^* \circ \lift{\id_{\O(Y)}}{\starred{\E_g}} = \starred{\E_g}$.
    It follows that $\mathfrak{A}_g \simeq \O(Z)$
    and that $g$ is uniformly envelopable.
\end{proof}

\section{Examples}

To conclude we collect a few toy examples for the purpose of building some basic intuition about the main definitions and 
of observing some of the basic phenomena that may occur.

Consider the functions
\[
    f \colon \R \to \R, 
    \; 
    f(x) 
    = 
    \begin{cases}
        -x &\text{if }x \leq 0,\\
         1 &\text{if }x > 0,
    \end{cases}
\]
and
\[
    g \colon \R \to \R, 
    \; 
    g(x) 
    = 
    \begin{cases}
        -x &\text{if }x < 0,\\
         1 &\text{if }x \geq 0.
    \end{cases}
\]

Then both $f \circ g$ and $g \circ f$ are equal to the constant function with value $1$.
Let $\Kb(\R)$ denote the lattice of compact subsets of $\R$, ordered by reverse inclusion, with a bottom element added.
The best continuous approximation of $f$ with values in $\Kb(\R)$ is 
\[
    F \colon \R \to \Kb(\R),
    \; 
    F(x) = 
    \begin{cases}
        \{-x\} &\text{if }x < 0,\\
        \{0,1\} &\text{if }x = 0,\\ 
        \{1\}   &\text{if }x > 0.
    \end{cases}
\]
The best continuous approximation $G \colon \R \to \Kb(\R)$ of $g$ with values in $\Kb(\R)$ is the same function: $G(x) = F(x)$ for all $x \in \R$.

The envelope $F$ is uniformly universal.
The envelope $G$ is uniformly $\mathcal{R}$-universal.
It is however not uniformly universal.
The open set $(0,+\infty)$ is a robust property of $g(0)$ that is not witnessed by $G$.

We have $G \bullet F(x) = \{1\}$ for all $x \in \R$ 
and $F \bullet G(x) = \{1\}$ for all $x \in \R$ with $x \neq 0$,
but $F \bullet G(0) = \{0,1\}$.
Hence, the envelope $G \bullet F$ is again uniformly universal, 
but 
$F \bullet G$ already fails to be the best approximation of $f \circ g$
in the lattice $\Kb(\R)$.

The least advice bundle $\mathfrak{A}_g$ of $g$ assigns to every open set $V \in \O(\R)$
a fibre $\mathfrak{A}_g(V)$ which is an effective injective space.
The fibres can be up to isomorphism described as follows: 
\[
    \mathfrak{A}_g(V) 
        \simeq 
        \begin{cases}
            \Sigma   &\text{if }1 \in V, 0 \notin V, \text{ and }(0,\varepsilon) \subseteq V\text{ for some }\varepsilon > 0,\\ 
            \{ \bot \} &\text{otherwise.}
        \end{cases}
\] 
Let $\R_<$ denote the space of real numbers with the Scott topology for the usual ordering.
Let $\R_>$ denote the space of real numbers with the Scott topology for the opposite ordering.
The space $\R_<$ is also called the space of \emph{lower reals}.
The space $\R_>$ is also called the space of \emph{upper reals}.
A reasonable computability structure on $\mathfrak{A}_g$ is obtained by identifying it with a quotient of the space 
$
    \Set{(V,\varepsilon) \in \O(Y) \times \R_<}
        {(0, \varepsilon) \subseteq V}.
$
We identify pairs $(V,\varepsilon)$ and $(V,\delta)$ as follows:
If 
$1 \in V$
and 
$0 \notin V$, 
then 
$(V,\varepsilon) \sim (V,\delta)$
if and only if the predicates
$\varepsilon > 0$
and 
$\delta > 0$
have the same truth value.
For all other $V$ we identify all $(V,\varepsilon)$ and $(V,\delta)$.
The resulting quotient space is easily seen to be homeomorphic to $\mathfrak{A}_g$.

Let $\pi_{\mathfrak{A}_g} \colon \mathfrak{A}_g \to \O(\R)$ denote the projection of the advice bundle $\mathfrak{A}_g$ onto $\O(\R)$. 
We can compute the greatest continuous lift 
$\lift{\pi_{\mathfrak{A}_g}}{\starred{F}} \colon \O(\R) \to \mathfrak{A}_g$
of $\starred{F}$ along $\pi_{\mathfrak{A}_g}$
explicitly: 
Given $V \in \O(\R)$, output the class 
$[(W, \varepsilon)] \in \mathfrak{A}_g$, where 
$W = \starred{F}(V)$,
$\varepsilon = 1$
if 
$1 \in V$,
and 
$\varepsilon = 0$
if $1 \notin V$.
Hence, the extra information that is needed to verify that 
$(0,+\infty)$ 
is a robust property of 
$g(0)$ 
can be provided by 
$\lift{\pi_{\mathfrak{A}_g}}{\starred{F}}$.
We obtain that 
$\starred{G} \bullet \starred{F} \colon \O(\R) \to \O(\R)$
is equal to $(f \circ g)^{*} = \intr{(f \circ g)}$.

It is worth noting that $\mathfrak{A}_g$ is not isomorphic to a space of the form $\O(Z)$.
It is not even distributive.
If we call an advice bundle $A$ \emph{spatial} if $A$ is isomorphic to $\O(Z)$ for some $Z$,
then we observe that there always exists a spatial advice bundle whose associated principal co-envelope is universal.
However, one can show that for the function $g$ there exists no least spatial advice bundle,
\emph{i.e.}, there exists no spatial advice bundle which is the least advice bundle among the spatial ones. 

As a second example we consider various types of limit operators on sequences of real numbers.
Let
$\operatorname{lim}_{[0,1]} \colon \subseteq [0,1]^{\N} \to [0,1]$
be the function which sends a convergent sequence in $[0,1]$ to its limit.
Let 
${\lim}_{\R} \colon \subseteq \R^{\N} \to \R$
be the function which sends a convergent sequence in $\R$ to its limit.
Let 
$\operatorname{Mlim}_{(-\infty,1]} \colon \subseteq (-\infty,1]^{\N} \to (-\infty,1]$
be the function which sends a bounded monotonically increasing sequence in $(-\infty, 1]$ to its limit.
Let 
$\operatorname{Mlim}_{\R} \colon \subseteq \R^{\N} \to \R$
be the function which sends a bounded monotonically increasing sequence in $\R$ to its limit.

All four functions are Weihrauch equivalent \cite[Theorem 8.9]{WeihrauchSurvey},
so that they have the same ``computational power''.
However, the amount of continuously obtainable information,
as encoded by the primary envelope,
is quite different for each function.

The functions 
${\lim}_{[0,1]}$ 
and 
$\operatorname{Mlim}_{(-\infty, 1]}$ 
are uniformly envelopable.
The primary envelope of 
${\lim}_{[0,1]}$
is the constant map 
$[0,1]^{\N} \to \K([0,1])$
which sends every point to $[0,1]$.
The primary envelope of 
$\operatorname{Mlim}_{(-\infty, 1]}$ 
sends 
$(x_n)_n$ to 
$[\lim x_n, 1] \in \Kb((-\infty, 1])$.

Neither 
${\lim}_{\R}$ 
nor 
$\operatorname{Mlim}_{\R}$
are uniformly envelopable.
The least advice bundle
$\mathfrak{A}_{{\lim}_\R}$
for ${\lim}_\R$ can be described as follows:
\[
    \mathfrak{A}_{{\lim}_\R} (U) 
    \simeq
    \begin{cases}
        \{\bot\}   &\text{if } U \neq \R\\
        \Sigma &\text{if } U = \R.
    \end{cases}
\]
This bundle is spatial, 
$\mathfrak{A}_{{\lim}_\R} \simeq \O(\R_{\bot})$.
The primary envelope can be represented spatially as 
\[
    F \colon \subseteq \R^{\N} \to \O^2(\R_{\bot}),
    \;
    F((x_n)_n) = \nu_{\R_{\bot}}(\bot).
\]

We carry out the computation of the least advice bundle of $\operatorname{Mlim}_{\R}$ in some detail,
following the construction in the proof of Theorem \ref{Theorem: least complete advice bundle}.
An open set $V \in \O(\R)$ is a robust property of 
$\operatorname{Mlim}_{\R}((x_n)_n)$
if and only if it contains the interval $[\lim_{n \to \infty} x_n, +\infty)$.
It follows that the principal 
$\O^2(\R)$-envelope 
of 
$\operatorname{Mlim}_{\R}$ 
is the map which sends every point to $\emptyset \in \O^2(\R)$. 
Picking up the notation of the proof of 
Theorem \ref{Theorem: least complete advice bundle},
let 
\[ 
    L_{\operatorname{Mlim}} =
        \Set{(U,V) \in \O(\dom (\operatorname{Mlim}_{\R})) \times \O(\R)}
            {U \subseteq \operatorname{Mlim}_{\R}^{-1}(V)}.
\]
Let 
$
    P_{\operatorname{Mlim}_{\R}} \colon L_{\operatorname{Mlim}_{\R}} \to L_{\operatorname{Mlim}_{\R}}
$ 
be the greatest continuous function such that 
$P_{\operatorname{Mlim}_{\R}}(U,V) = (U',V)$.
This function is explicitly computable.
Using that the eventually constant monotone sequences are dense in $\dom(\operatorname{Mlim}_{\R})$,
it is not difficult to see that given an open set 
$U$ in $\O(\dom(\operatorname{Mlim}_{\R}))$
one can compute the number 
\[ 
    I(U) = \inf\Set{\lim_{n \to \infty} x_n}
                    {(x_n)_n \in U}
            \in [-\infty,+\infty]_{>}
\]
as an upper real.
It follows that we have
\[
    P_{\operatorname{Mlim}_{\R}}(U,V) = 
        \left(
            \Set{(x_n)_n \in \dom (\operatorname{Mlim}_{\R})}
             {\lim_{n \to \infty} x_n > I(U)}
        ,V
        \right).
\] 
For an open set $V \in \O(\R)$,
let 
$r(V) = \inf\Set{s \in \R}{(s, +\infty) \subseteq V}$.
Then for all $V \in \O(\R)$ we have by the above:
\[ 
    \mathfrak{A}_{\operatorname{Mlim}_{\R}}(V) 
    \simeq 
    \Set{x \in [-\infty,+\infty]_{>}}{x > r(V)}.
\]
We can hence identify 
$\mathfrak{A}_{\operatorname{Mlim}_{\R}}$
with a subspace of 
$\O(\R) \times [-\infty, +\infty]_{>}$.
Under this identification, the primary envelope of $\operatorname{Mlim}_{\R}$ is:
\[
    F \colon \subseteq \R^{\N} \to \O(\mathfrak{A}_{\operatorname{Mlim}_{\R}}),
    \; 
    F((x_n)_n)
    =
    \Set{(V, s) \in \mathfrak{A}_{\operatorname{Mlim}_{\R}}}
        {\lim_{n \to \infty} x_n > s}.
\]

\section*{Acknowledgements.}
\includegraphics[scale = 0.05]{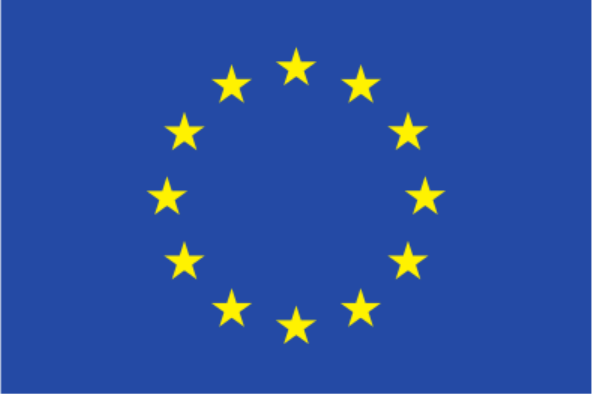}
This project has received funding from the European Union's Horizon 2020 research and innovation programme under the Marie Sk\l{}odowska-Curie grant agreement No 731143.
I would like to thank an anonymous referee for many detailed suggestions and corrections.
In particular, the referee pointed out that an additional assumption that was made in an earlier version of
Theorem \ref{Theorem: spaces X such that all f with domain X are uniformly envelopable}
was superfluous, leading to a substantial strengthening of the result.

\bibliographystyle{alphaurl}
\bibliography{uniform}
\nocite{*}

\appendix
\section{Proof of Proposition \ref{Proposition: examples where exponential has compact-open topology}}
\label{Appendix: Proof of compact-open proposition}

To finish the proof of Proposition \ref{Proposition: examples where exponential has compact-open topology}, it suffices to establish the following result:

\begin{prop}
    Let $L$ be an $\omega$-continuous lattice.
    Let $Z$ be a Polish space.
    Then the Scott topology on $L^Z$ coincides with the compact-open topology.
\end{prop}
\begin{proof}
    We closely follow the proof of \cite[Theorem 4.1]{Consonant}, which establishes the result in the special case where $L = \Sigma$.
    The Scott topology is finer than the compact-open topology, so that it suffices to show that every Scott-open set is open in the compact-open topology.
    
    Thus, let $\mathcal{U} \subseteq L^Z$ be Scott-open.
    Let $\varphi \in \mathcal{U}$.
    Our goal is to show that $\varphi \in N \subseteq \mathcal{U}$,  
    where $N$ is an open set in the compact-open topology.
    
    Since $L$ is $\omega$-continuous, it has a countable basis $(b_n)_n$.
    Choose a countable basis $(U_n)_n$ of the topology on $Z$.
    Since $L$ is a complete lattice, $L^Z$ is again a complete lattice.
    Suprema in $L^Z$ are given pointwise. 

    For $(j,k) \in \N^2$, let 
    \[
        \chi_{j,k} \colon Z \to L,
        \;
        \chi_{j, k} (z) = 
            \begin{cases}
                b_k &\text{if }z \in U_j \\
                \bot &\text{otherwise.}
            \end{cases}
    \]
    Let 
    \[
        S = 
                \Set
                    {
                        \chi_{j,k}
                    }
                    {
                        j \in \N,
                        \varphi(z) \gg b_k 
                        \; 
                        \text{ for all }z \in U_j
                    }.
    \]
    Then 
    $
        \varphi = \sup S              
    $.
    Since $\mathcal{U}$ is Scott-open and contains $\varphi$, 
    it contains the supremum over a finite subset of $S$. 
    
    It follows that
    there exist non-empty open sets 
    $U_{0, 1},\dots,U_{0, s} \in \O(Z)$
    and lattice elements 
    $
    \ell_1
    \dots,
    \ell_s
    \in 
    L
    $
    such that 
    $
        \varphi(z) \gg \ell_j
    $
    for all $z \in U_{0, j}$
    and
    $\mathcal{U}$ contains all $\psi \colon L \to Z$ satisfying 
    $
        \psi(z) \geq \ell_j
    $
    for all $z \in U_{0,j}$.

    Fix a complete metric that generates the topology on $Z$.
    For $n \in \N$, let $\mathcal{B}_n$ denote the set of all balls of radius $2^{-n}$ with respect to that metric.
    For an open set $W \in \O(Z)$ and $j \in \{1,\dots,s\}$, let 
    $\psi_{W,j} \colon Z \to L$
    be the function which sends all elements of $W$ to $\ell_j$ and all elements of $Z \setminus W$ to $\bot \in L$.
    
    Each set $U_{0, j}$ is the union of all open sets $W \in \O(Z)$ satisfying 
    $\clos{W} \subseteq U_{0, j} \cap B$,
    where $B \in \mathcal{B}_1$.

    Let 
    \[
        A_0 = 
                \Set{\psi_{W,j}}
                {
                j \in \{1,\dots,s\},
                W \in \O(Z),
                \exists B \in \mathcal{B}_1. \left(\clos{W} \subseteq U_{0, j} \cap B\right)
                }.
    \]
    Then, by construction, the supremum of $A_0$ is contained in $U$.
    Since $U$ is Scott-open, there exists a finite subset of $A_0$ whose supremum is contained in $U$.

    Thus, we obtain finitely many non-empty open sets 
    $
    U_{1, 1},
    \dots,
    U_{1, s} 
    \in 
    \O(Z)
    $
    such that 
    $\clos{U_{1,j}} \subseteq U_{0,j}$,
    the union 
    $\bigcup_{j = 1}^s U_{1,j}$
    is covered by finitely many elements of $\mathcal{B}_1$,
    and
    $U$ 
    contains all $\psi \colon Z \to L$ such that for all 
    $j = 1, \dots, s$
    and all 
    $z \in U_{1,j}$ 
    we have 
    $\psi(z) \geq \ell_j$.

    By an analogous argument we obtain for all $n \in \N$ a collection of finitely many non-empty open sets 
    $
    U_{n, 1},
    \dots,
    U_{n, s}
    \in 
    \O(Z)
    $,
    such that 
    $\clos{U_{n + 1, j}} \subseteq U_{n, j}$,
    the union 
    $\bigcup_{j = 1}^s U_{n,j}$
    is covered by finitely many elements of $\mathcal{B}_n$,
    and
    $U$ 
    contains all $\psi \colon Z \to L$ such that for all 
    $j = 1, \dots, s$
    and all 
    $z \in U_{n,j}$ 
    we have 
    $\psi(z) \geq \ell_j$.

    For $j = 1, \dots, s$, consider the set 
    \[
        K_j =  
            \bigcap_{n \in \N} 
                    \clos{U_{n, j}}.
    \]
    By K\H{o}nig's lemma and completeness of $Z$, the set $K_j$ is non-empty.
    It is closed and totally bounded by construction, and hence compact.
    For $j = 1,\dots,s$, let 
    $V_j = \Set{x \in L}{x \gg \ell_j}$.
    
    Let 
    \[
        N = \Set{
                    \psi \colon Z \to L
                }
                {
                    \forall j \in \{1,\dots,s\}. 
                        \psi(K_j) \subseteq V_j
                }.
    \]
    Then $N$ is an open neighbourhood of $\varphi$ in the compact-open topology by construction.
    If $\psi \in N$, then for all $j = 1,\dots,s$ there exists $n \in \N$ such that 
    $\psi$ sends $U_{n, j}$ to $V_j$,
    so that $\psi \in \mathcal{U}$.
    This proves the claim.
\end{proof}

\end{document}